\documentclass[3p,twocolumn,fleqn]{elsarticle}

\usepackage[numbers]{natbib}

\usepackage{subcaption}
\usepackage{geometry}
\usepackage{graphicx}
\graphicspath{ {./Figures/} }

\usepackage{txfonts}
\usepackage{amsfonts}
\usepackage{amssymb}

\usepackage{gensymb}
\usepackage{siunitx}

\usepackage{tikz}
\usetikzlibrary{patterns}
\usetikzlibrary{plotmarks}
\usetikzlibrary{decorations}

\usepackage[breaklinks=true]{hyperref}

\usepackage{cleveref}
\crefname{figure}{Fig.\@}{Figs.\@}
\crefname{equation}{Eq.\@}{Eqs.\@}
\crefname{section}{Sec.\@}{Secs.\@}
\crefname{table}{Table}{Tables}

\usepackage{color}
\usepackage[parfill]{parskip}

\begin{document}

\title{The DarkLight Experiment at TRIUMF}  



%







\affiliation[1]{
    organisation={Arab American University},
    addressline={13 Zababdeh},
    city={Jenin},
    country={Palestine}
}

\affiliation[2]{
    organisation={Arizona State University},
    addressline={1151 South Forest Avenue},
    city={Tempe},
    state={AZ}, statesep={},
    postcode={85287},
    country={USA}
}

\affiliation[3]{
    organisation={Center for Frontiers in Nuclear Science, Stony Brook University},
    city={Stony Brook},
    state={NY}, statesep={},
    postcode={11794},
    country={USA}
}

\affiliation[4]{
    organisation={Hampton University},
    addressline={100 East Queen Street},
    city={Hampton},
    state={VA}, statesep={},
    postcode={23668},
    country={USA}
}

\affiliation[5]{
    organisation={Johannes Gutenburg-Universit{\"a}t Mainz},
    addressline={Saarstra{\ss}e 21},
    city={Mainz}, citysep={},
    postcode={55122},
    country={Germany}
}

\affiliation[6]{
    organization={Massachusetts Institute of Technology, Laboratory for Nuclear Science},
    addressline={77 Massachusetts Avenue}, 
    city={Cambridge},
    state={MA}, statesep={},
    postcode={02139}, 
    country={USA}
}

\affiliation[7]{
    organisation={Saint Mary's University},
    addressline={923 Robie Street},
    city={Halifax},
    state={NS}, statesep={},
    postcode={B3H 3C3},
    country={Canada}
}

\affiliation[8]{
    organisation={Thomas Jefferson National Accelerator Facility},
    addressline={12000 Jefferson Avenue},
    city={Newport News},
    state={VA}, statesep={},
    postcode={23606},
    country={USA}
}

\affiliation[9]{
    organisation={TRIUMF},
    addressline={4004 Wesbrook Mall},
    city={Vancouver},
    state={BC}, statesep={},
    postcode={V6T 2A3},
    country={Canada}
}

\affiliation[10]{
    organisation={University of British Columbia},
    addressline={6224 Agricultural Road},
    city={Vancouver},
    state={BC}, statesep={},
    postcode={V6T 1Z1},
    country={Canada}
}
\affiliation[11]{
    organisation={University of Manitoba},
    addressline={66 Chancellors Circle},
    city={Winnipeg},
    state={MB}, statesep={},
    postcode={R3T 2N2},
    country={Canada}
}
\affiliation[12]{
    organisation={University of Victoria},
    addressline={3800 Finnerty Road},
    city={Victoria},
    state={BC}, statesep={},
    postcode={V8P 5C2},
    country={Canada}
}

\affiliation[13]{
    organisation={University of Winnipeg},
    addressline={515 Portage Avenue},
    city={Winnipeg},
    state={MB}, statesep={},
    postcode={R3B 2E9},
    country={Canada}
}

\affiliation[14]{
    organisation={University of Zagreb},
    addressline={Bijenička c. 32},
    city={Zagreb}, citysep={},
    postcode={10000},
    country={Croatia}
}

\author[2]{R.~Alarcon}
\author[3]{D.~Almonte}
\author[1,6]{M.I.~Alstaty}
\author[9]{J.~Azzi}
\author[9]{R.~Baartman}
\author[3]{J.C.~Bernauer}
\author[3]{X.~Braun}
\author[9]{D.~Ciarniello}
\author[3]{C.~Clabeaux}
\author[6,3]{E.W.~Cline\corref{cor1}}
    \ead{ewcline@mit.edu}  
\author[9]{E.~Corey}
\author[3]{R.~Corliss}
\author[11]{W.~Deconinck}
\author[3]{A.~Deshpande}
\author[8]{J.~Dilling}
\author[6]{J.~Dodge}
\author[3]{D.H.~Dongwi}
\author[3]{C.~Ekeman}
\author[9,10]{E.A.~Fielding}
\author[6]{S.~Frantzen}
\author[14]{I.~Friščić}
\author[10]{G.~Gelinas}
\author[9]{S.~Gelinas}
\author[9]{K.~Gill}
\author[9]{A.~Sabzevari~Gonzalez}
\author[3]{S.~Gupte}
\author[6]{D.K.~Hasell}
\author[10]{M.~Hasinoff}
\author[9]{B.~Humphries}
\author[6]{E.~Ihloff}
\author[3]{N.~Jackson}
\author[4]{D.~Jayakodige}
\author[7,9]{R.~Kanungo}
\author[6]{J.E.~Kelsey}
\author[9,12]{O.~Kester}
\author[4]{M.~Kohl}
\author[9]{H.W.~Koay}
\author[9,12]{R.~Laxdal}
\author[3]{L.~Levack}
\author[6]{X.~Li}
\author[3]{W.~Lin}
\author[3]{C.~Ma}
\author[9,12]{A.~Mahon}
\author[5]{H.~Merkel}
\author[13]{J.W.~Martin}
\author[9]{L.~Miller\corref{cor1}}
    \ead{lmiller@triumf.ca}
\author[6]{R.~Milner}
\author[9]{K.~Olchanski}
\author[9,10]{K.~Pachal}
\author[9,12]{T.~Planche}
\author[9]{D.~Preddy}
\author[9]{S.D.~R{\"a}del}
\author[2]{G.~Randall}
\author[9]{J.~Redinger}
\author[10]{D.A.~Reiter}
\author[4]{R.~Richards}
\author[9]{B.~Scully}
\author[9]{S.~Shapiro}
\author[4]{M.~Suresh}
\author[6]{C.~Vidal}
\author[9,10]{S.~Wang}
\author[6]{H.~Witte}
\author[9,10]{H.~Yang}
\author[9]{S.~Yen}

\cortext[cor1]{Corresponding author}



\begin{abstract}
This paper reports on the commissioning of the DarkLight experiment at the ARIEL superconducting electron linear accelerator located at TRIUMF in Vancouver, Canada. DarkLight was designed to utilize the ARIEL electron beam to search for a new boson, $A^\prime$, which preferentially couples to leptons, with a mass in the range 13--17~MeV/$c^2$. Such a boson could be produced via the $e^-~X\rightarrow e^-~X~A^\prime$ process, and the resulting $A^\prime\rightarrow e^+~e^-$ decay used to reconstruct the invariant mass of the $A^\prime$.   
The $e^+e^-$ pair is detected using a pair of magnetic spectrometers.
Each spectrometer is instrumented with two triple GEM detectors and a scintillating strip hodoscope to trigger the event readout. Electron beam energies between 10 and 30~MeV were used to commission the experiment with a \SI{1}{\micro\metre} thick carbon target. With both spectrometer polarities set to select electrons, elastic and M{\o}ller scattering were used to examine detector performance and compare with predicted distributions from Monte Carlo simulations.
Planned upgrades to the e-linac will allow searches to be made with beam energies up to 50~MeV using a \SI{1}{\micro\metre} thick tantalum target.
\end{abstract}





\maketitle
{  
\hypersetup{hidelinks}
\tableofcontents
}


\section{Introduction}

Notwithstanding the enormous success of the Standard Model, it is not a complete description of nature. In particular, it does not provide an explanation for the origin and nature of dark matter, despite it constituting around 85\% of the mass in the universe. The search for dark matter and/or physics beyond the Standard Model is therefore one of the most important endeavours in physics today.  

The ATOMKI group's reports of anomalies in the electromagnetic decays of excited states of $^4$He, $^8$Be, and $^{12}$C nuclei \cite{Krasznahorkay:2021joi, Krasznahorkay:2015iga, Krasznahorkay:2022pxs, Krasznahorkay:2023sax} have been the subject of much theoretical speculation~\cite{PhysRevLett.117.071803,Mommers:2024qzy,serao2026x17anomalyexperimentalevidence}. One possible explanation posits the existence of a light, neutral boson, $A^\prime$, with a mass of around 17~MeV/$c^2$, that couples to leptons~\cite{PhysRevLett.128.091802}. Such a particle could be the carrier of a new ``fifth force" in the Standard Model that couples to a dark sector. In light of this, numerous experiments have sought to verify or refute the existence of this anomaly but they have so far proven inconclusive~\cite{PADME:2025dla,MEGII:2024urz}.

\begin{figure}
{\includegraphics[width=0.47\textwidth]{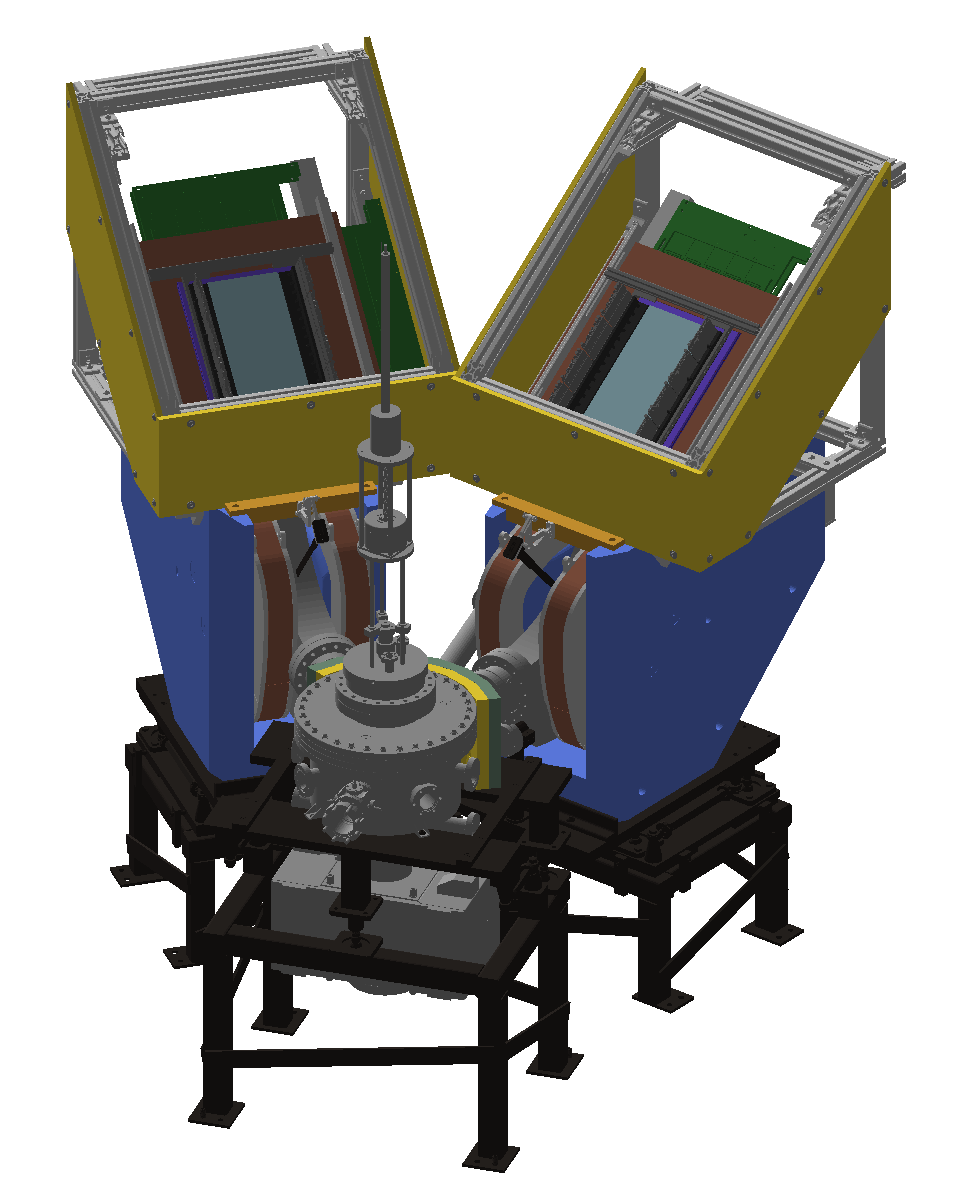}
\caption{\label{DL_Layout}
Geant4 representation of the experimental layout showing the scattering chamber with target ladder, two magnetic spectrometers each with two triple GEM detectors, and a scintillating strip trigger detector.}
}
\end{figure}

Motivated by this hypothetical particle, the DarkLight experiment, originally conceived to take data at the Jefferson Lab Free Electron Laser~\cite{PhysRevLett.111.164801,LEE201946}, was redesigned~\cite{DarkLight:2022uji} to use the electron beam from the Advanced Rare IsotopE Laboratory (ARIEL) superconducting electron linear accelerator (e-linac) at TRIUMF. The experiment will search for new bosons with masses in the range 13--17~MeV/$c^2$.  The experiment uses the process $e^-X\rightarrow e^-XA^\prime(\rightarrow e^+e^-)$ to look for a resonant excess of $e^+e^-$ pairs at the invariant mass of the $A^\prime$.
The decay leptons are detected by a pair of magnetic dipole spectrometers arranged asymmetrically around a fixed foil target placed in the beam of the e-linac (see~\cref{DL_Layout}).

The e-linac has a relatively low beam energy compared to the hypothesized new boson masses. Thus, such a new particle would be produced with limited boost and the resulting forward-going decay leptons would be well-separated in angle.
In the first (commissioning) phase of the experiment described in this paper, an asymmetrical arrangement of the two magnetic spectrometers shown in \cref{DL_Layout}, with one at a 36\degree~angle and the other at a 20\degree~angle, was chosen as optimal for a new 13~MeV/$c^2$ boson.  A thin \SI{1}{\micro\metre} carbon foil target was used to minimize multiple scattering in the target. 

For the commissioning run of 2025-2026, electron beam energies in the range 10--30~MeV were used. This allowed for the study of the electron beam parameters and operation, in addition to allowing for detector calibration and measures of performance. 
Details on all aspects of the accelerator, experimental hardware, detectors, data acquisition, simulation, and analysis will be presented in this paper.

Another run at 10--30~MeV is planned in 2027 to further commission the experiment and collect a higher-statistics dataset. Following this, the linac will be upgraded to enable 50~MeV operation and a subsequent experimental run will use a \SI{1}{\micro\metre} thick tantalum target to continue the search for the $A^\prime$ around 17~MeV/$c^2$.

\section{TRIUMF's Electron Linear Accelerator}

The TRIUMF e-linac can currently produce an electron beam of up to 30~MeV in energy 
~\cite{raedel:erl2019-wepnec01}. Its primary purpose is as a driver for ARIEL, where the e-linac delivers electrons to a photo-converter target station for the production of neutron-rich rare isotope beams via photo-fission.

The e-linac has a maximum bunch frequency of 650~MHz and can be operated in either a continuous wave (CW) or pulsed mode. For the 2025-2026 DarkLight commissioning run, it was operated in pulsed mode with a variety of duty factors. CW operation is planned for future runs. The e-linac supports an average beam current up to \SI{300}{\micro\ampere}, although most commissioning data was taken with a CW-equivalent beam current of less than 10~nA.  

A competitive search in the 17~MeV/$c^2$ mass range requires a beam energy increase to 50~MeV. To achieve this, the upgrades to the e-linac will need to include a second accelerating cryomodule downstream of the current cryomodule. To reach the highest possible luminosity a new beamline and beamdump specifically for the DarkLight experiment would be desirable. To allow parallel running with the ARIEL facility a septum magnet and RF deflector could be installed to allow for simultaneous 50~MeV beam delivery with alternating bunches sent to DarkLight or ARIEL.

\subsection{Beam Optics}
\begin{figure}
{\centering 
\includegraphics[width=\linewidth]{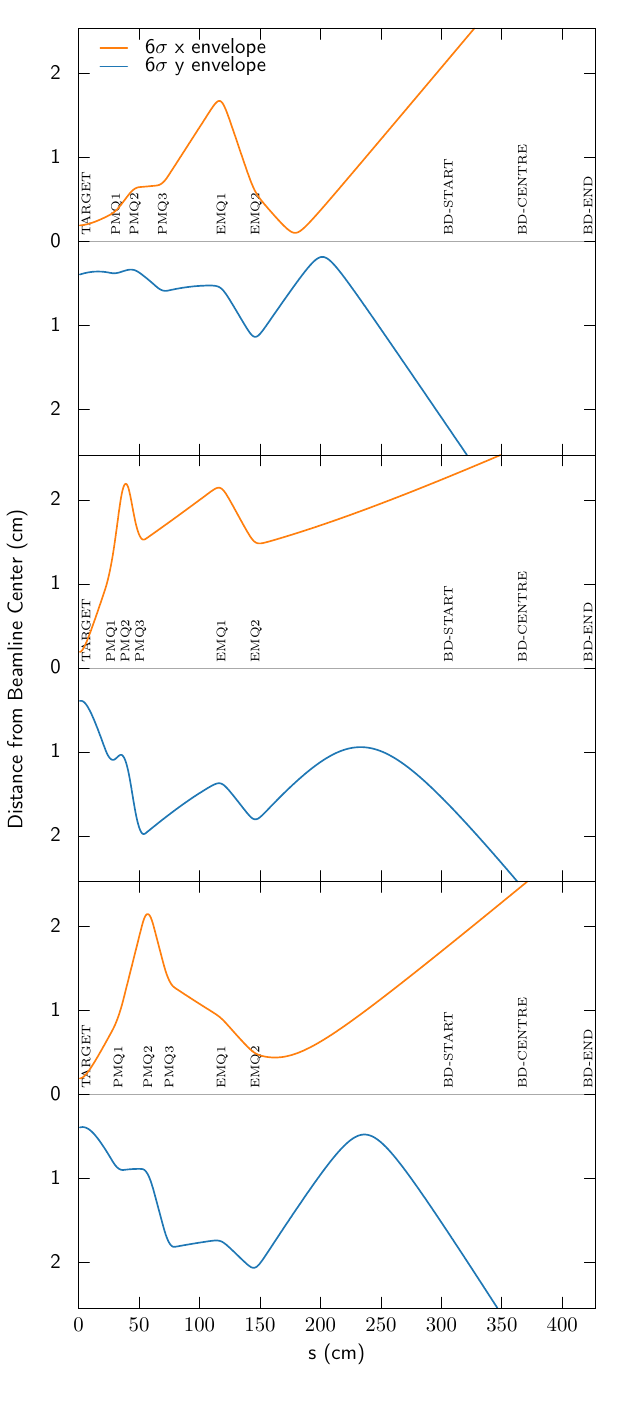}
\caption{\label{beamprofile}The $6\sigma$ $x$ and $y$ beam envelopes from {\sc transoptr} shown for \textit{Top:} A 30~MeV beam through a \SI{1}{\micro\metre} thick carbon target. \textit{Middle:} A 30~MeV beam through a \SI{1}{\micro\metre} thick tantalum target. \textit{Bottom:} A 50~MeV beam through a \SI{1}{\micro\metre} thick tantalum target. The text in the figure denotes the positions of the DarkLight target, three permanent quadrupoles, two electromagnetic quadrupoles, and the beginning, centre, and end of the beam dump respectively. The optics were optimized to keep the $6\sigma$ envelopes within one inch (2.54~cm) from the center of the beamline before the envelope passes the entrance to the beam dump. The vertical scale of each panel represents the 2.54~cm limit.}}
\end{figure}

\begin{figure}
{\centering
\includegraphics[width=\linewidth]
{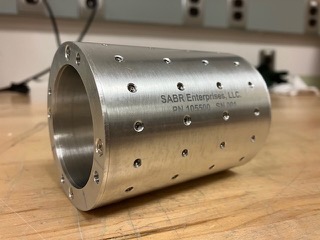}}
\caption{\label{PMQ}One of the permanent magnet quadrupoles.}
\end{figure}

A beam optics design was optimised at TRIUMF to ensure that the electron beam was transported safely to the beam dump. The benchmark used for this optimisation was that the 6$\sigma$ beam envelope must remain inside the two-inch diameter beampipe at every point between the target and the dump. Because of the significant scattering occurring in the target, meeting this benchmark criterion required the use of five quadrupoles in this short section of beamline. The optics model for the 30 MeV test experiment using a 1~$\mu$m carbon (tantalum) target can be seen in the top (middle) panel of Fig.~\ref{beamprofile}, which shows the 6$\sigma$ beam profile. The optics design selected for the 50~MeV experiment with a 1~$\mu$m thick tantalum target is shown in the bottom panel of~\cref{beamprofile}.  The target is located at the {\sc target} point, $\text{s}=0$~cm, and the midpoint of the beam dump is at {\sc bd-centre}, $\text{s}\approx375$~cm. The beam optics of the experiment was developed using the {\sc transoptr} code~\cite{TRI-DN-24-15}, and incorporated into the FLUKA~\cite{FLUKA0,FLUKA1,FLUKA2} description of the experiment~\cite{mahon2026analyticdescriptionssoftedgequadrupoles}.

In order to handle the divergence of the beam after it passes through the target foil three permanent magnet quadrupoles (PMQ) (see~\cref{PMQ})
\footnote{Sabr Enterprises LLC, North Andover, MA 01845, USA}
were required immediately after the scattering chamber. These were chosen for their compact size, which allows the spectrometer dipoles to approach the beamline as closely as possible. 
The magnetic material used was Sm$_2$Co$_{17}$ and each quadrupole consists of 16 trapezoidal shaped magnets contained in an aluminium housing.
Two solutions for the triplet PMQs were needed:
\begin{enumerate}
    \item with an integrated field strength of 0.3~T for each PMQ allowed beam transport solutions for 10--30~MeV beams on the \SI{1}{\micro\metre} carbon and tantalum targets, and
    \item with an integrated field strength of 0.9~T for two PMQs, and 0.55~T for the final PMQ, for 20--50~MeV beams on the \SI{1}{\micro\metre} tantalum target.
\end{enumerate}

A pair of standard TRIUMF electromagnetic quadrupoles are used for the final focus of the beam into the centre of the beamdump.  The entrance to the beam dump is near s$\approx$300~cm and the beam dump mid-point is around s$\approx$375~cm.

\subsection{Beam Interaction with the Target}\label{sec:BeamTarget}

The beam interaction with the target is significant in spite of it being only \SI{1}{\micro\metre} thick. 
Numerous effects must be considered:
\begin{itemize}
    \item energy loss in the foil leading to heat that must be dissipated or risk melting the foil,
    \item multiple scattering that could generate background,
    \item photons produced via Bremsstrahlung radiation that do not decay promptly and therefore contribute to background,
    \item neutron production that could damage nearby electronics.
\end{itemize}
Calculations made using the FLUKA code show the gamma radiation patterns and neutron fluxes pose a potential hazard to the detectors used for the experiment. Shielding was designed and implemented to mitigate the radiation exposure of sensitive electronics, see Secs.~\ref{sec:target_shielding} and~\ref{sec:detector_shielding}.

\section{DarkLight Experiment}

\begin{figure}
{\centering
\includegraphics[width=1.2\linewidth,angle=-90]{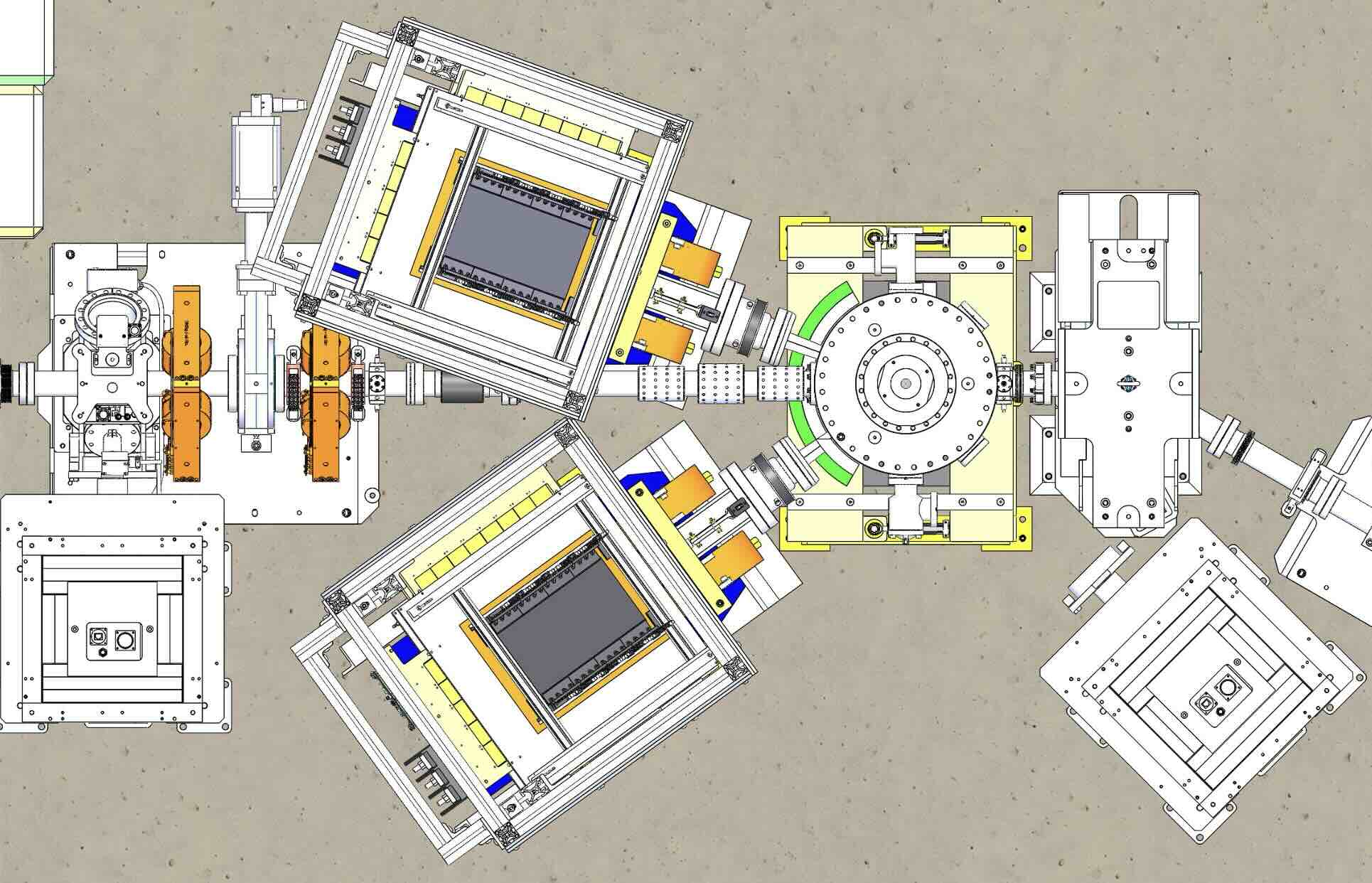}
\caption{\label{DLtop}DarkLight experiment viewed from above and slightly upstream of the experiment. The circle is the target chamber, continuing vertically up the figure are the three permanent beamline quadrupoles, the two spectrometer arms, and the two electro-magnetic quadrupoles.}}
\end{figure}

A top view of the CAD representation of the DarkLight experiment is given in~\cref{DLtop} showing the primary components.
Near the bottom is the last beamline dipole that bends the electron beam into the scattering chamber and onto the target.  The target ladder sits in the center of the scattering chamber. The two magnetic dipole spectrometers are shown to the right at $20\degree$ for the positron arm and to the left at $36\degree$ for the electron arm.  Also visible is the downstream beamline with the triplet of permanent magnet quadrupoles and the two TRIUMF quadrupoles that focus the beam into the beam dump (not shown).  On top of each spectrometer are two triple Gas-Electron Multiplier (GEM) detectors and a scintillating strip hodoscope.  The following sections will describe each component in more detail.

\subsection{Scattering Chamber and Target System}

\begin{figure}{\centering
\includegraphics[width=0.47\textwidth]{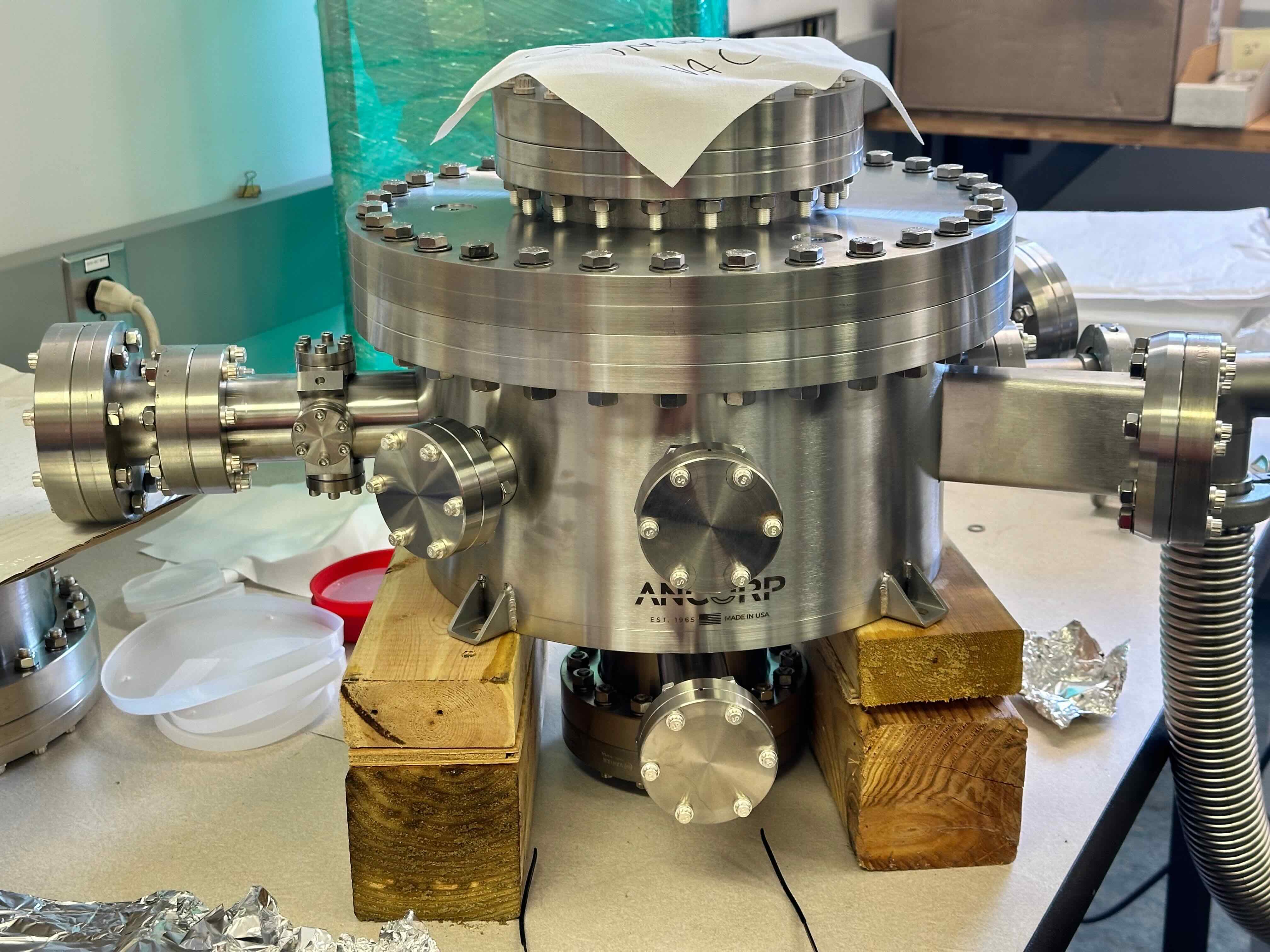}
\caption{\label{Scatch}DarkLight scattering chamber.}}
\end{figure}

The DarkLight scattering chamber was designed at MIT-Bates and manufactured by Ancorp Corporation\footnote{Ancorp Corporation, Williston, FL 32696 USA}. Figure \ref{Scatch} shows the scattering chamber being vacuum tested at MIT-Bates.

In normal operation, the electron beam enters via the flange located on the left of the figure, traverses the scattering chamber, and continues downstream via the centre beampipe on the right of the figure (connected to a flexible pumping tube in the picture). Also on the right, the visible rectangular tube  couples to the vacuum chamber of the $20\degree$ positron spectrometer. On the opposite side of the downstream beampipe, a similar coupling exists for the $36\degree$ electron spectrometer.
The large flange on top of the scattering chamber accepts the target ladder system. The flange on the bottom of the scattering chamber couples to an ion pump, and the flange perpendicular to and just above it was used to house a non-evaporable getter (NEG) pump. Together, these two pumps maintain the vacuum in the scattering chamber, the two spectrometer vacuum chambers and the upstream and downstream beamlines.
The small port with flange at $90\degree$ was used to monitor the system vacuum using a dual hot filament ion gauge and a cold cathode vacuum gauge.  A similar port on the opposite side of the scattering chamber (see \cref{DLtop}) has a manual valve and a tee connection that can be used for the initial rough pumping of the system or to vent the system to dry nitrogen. There are two small ports facing backwards at $\pm45\degree$.  The one on the left side was  equipped with a borosilicate glass window from Ancorp to permit optical viewing of the beamspot on the BeO screen. The port on the right side was similarly equipped with a zinc selenide infra-red compatible window to monitor the temperature of the target foil.

The scattering chamber with the target ladder, ion pump, and other fittings is supported by a welded stand and an adjustable six-strut system that can be used to position the ensemble into the proper position.  Fiducial targets on the scattering chamber can be used with a survey system to verify the position.

\subsubsection{Target Shielding}\label{sec:target_shielding}

As previously discussed, background from multiple-scattered beam electrons, Bremsstrahlung, and neutron production in the target is a serious concern both as a source of background in the detectors but also as a radiation hazard. A detailed FLUKA model of the experiment was produced and it was determined that 2.5~cm of lead and 2.5~cm of borated polyethylene should be placed around the scattering chamber downstream of the target. Further shielding around the detector stack is also necessary; see Sec.~\ref{sec:detector_shielding}.

\subsubsection{Target Ladder System}\label{sec:Target}

\begin{figure}[!htb]{\centering
\includegraphics[width=0.47\textwidth]{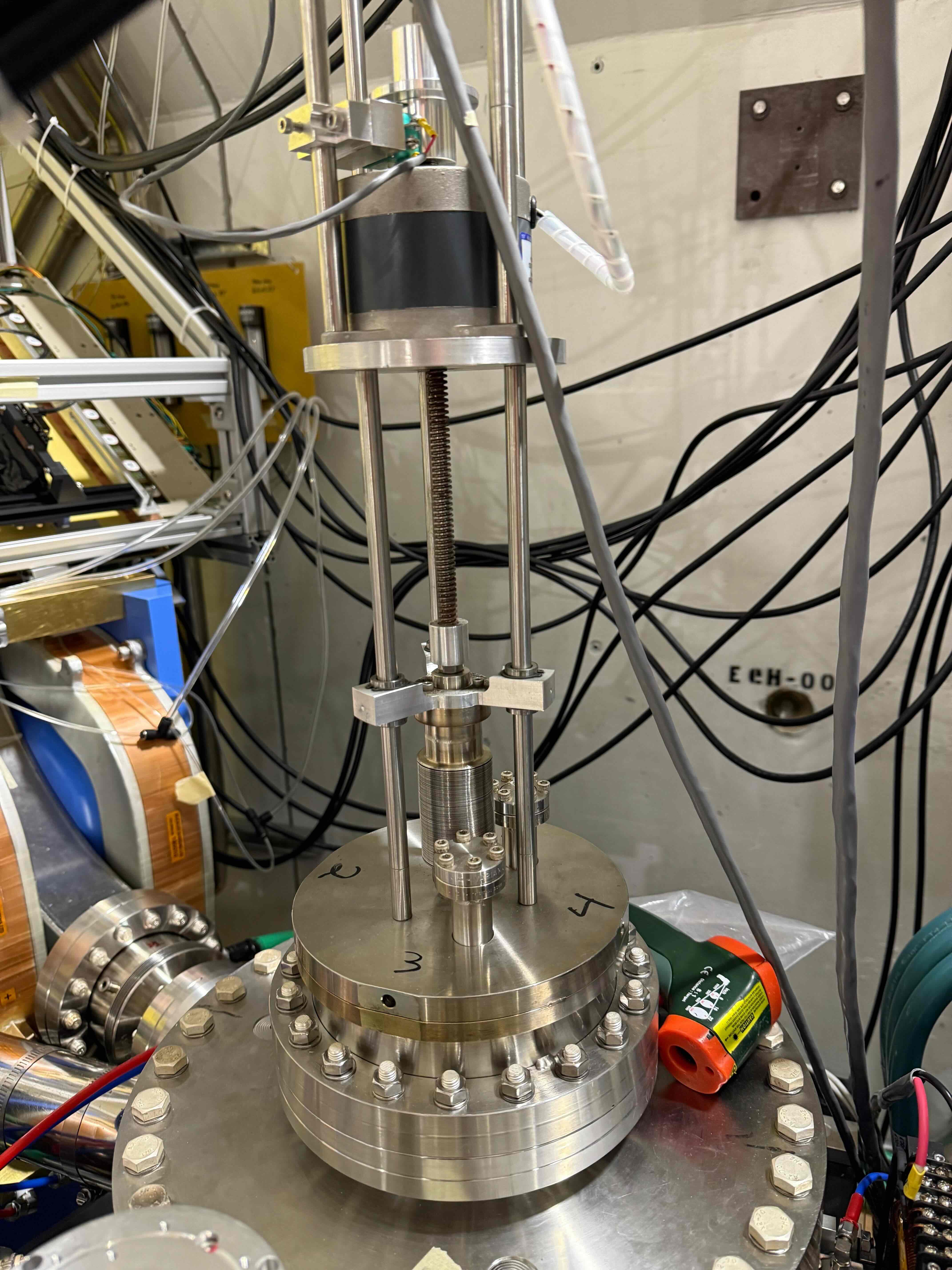}
\caption{\label{targlad}DarkLight target ladder system.}}
\end{figure}

On top of the scattering chamber, the target ladder system (see~\cref{targlad}) allows for the carbon or tantalum foils, or the BeO screen to be inserted into the electron beam's path. The ladder can also be driven out of the beam path entirely, for beam diagnostic measurements.  The target ladder is driven in and out of the scattering chamber using a Slo-Syn motor and the position is determined using a linearly variable resistor with remote readout. 

\subsection{Magnetic Spectrometers}\label{sec:magnet_spectrometers}

\begin{figure}[!htb]{\centering
\includegraphics[width=0.47\textwidth]{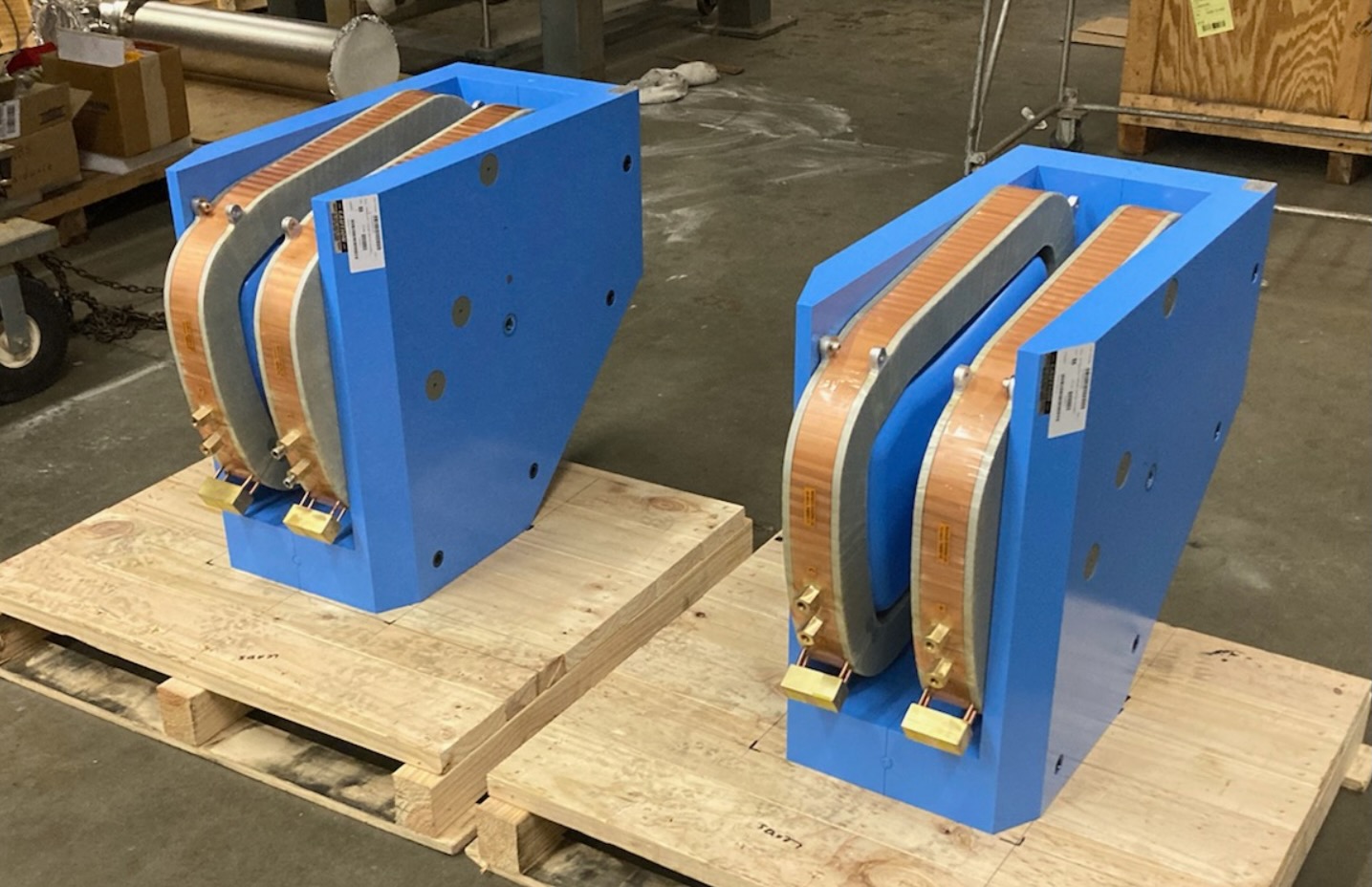}
\caption{\label{DLMag}DarkLight dipole magnets at MIT-Bates.}}
\end{figure}

The experiment uses two dipole spectrometers with identical magnetic characteristics.  The design originated from Mainz and was then refined at MIT-Bates to simplify and match material available from the manufacturer, Buckley Systems Limited\footnote{Buckley Systems Limited, Mt Wellington, Auckland 1060, New Zealand}.

The spectrometer design provides point-to-parallel focusing in the non-bend plane to get better resolution in the non-bend plane angle.

For the $20^{\circ}$ ($36^{\circ}$) spectrometer, the solid angle acceptance is restricted to 2.813 (4.834) msr, determined by tungsten collimators located before the entrance to each dipole and positioned 358.84~cm from the target. 
Each collimator permits angular acceptances of $\pm1.35^{\circ}$ in-plane and $\pm5.0^{\circ}$ out of plane.
These values were chosen to prevent particles from the target striking the inside surfaces of the spectrometer vacuum chambers and producing background in the detectors.

The momentum acceptance is $\pm20\%$ for both spectrometers.  Specifications are presented in Table~\ref{tabdesign}. 
\begin{table}[!htb]
\caption{\label{tabdesign}Design parameters for the spectrometers.}
\begin{tabular}{@{}rr@{}}  
In-plane acceptance & $\pm1.35\degree$\\
Out-of-plane acceptance & $\pm5.0\degree$\\
Momentum acceptance & $\pm20\,\%$ \\
Minimum central angle    & $16\degree$ \\
Nominal momentum of 30~MeV/$c$&192 A\\
Dipole field & 0.3485~T \\
Nominal bend radius & 30~cm\\
Pole gap & 8~cm\\
\end{tabular}
\end{table}

The spectrometers are rated for 225~A with 60 turns per coil for $N I = 13,500$ Amp-turns per coil. Because the power supplies for the magnets are located 30~m away from the magnets, 8~kW power supplies are required capable of providing 32~V and 250~A.  The coils are sandwiched between two aluminium plates that are cooled by circulating water. 

The two spectrometers have the same design but are operated at different currents and polarity to produce the desired magnetic fields for the scattered particle being detected. They are conventional iron-core magnets with simple, planar coils. The magnet design and pole face rotations were optimized for a \SI{0.5}{\metre} distance from target to spectrometer entrance and for post-magnet trajectories suitable for tracking with a pair of GEM detectors.

Each spectrometer magnet weighs about 950~kg.  The magnets have full fiducialization to allow for laser alignment and a six-strut mechanical support system to allow for alignment to an accuracy of around \SI{200}{\micro\metre}.

The CAD design of the dipole magnet was incorporated into Ansys Maxwell\footnote{Ansys inc. Southpointe, Canonsburg, PA 15317 USA} 3D to generate a simulated 3D field map which was used for trajectory simulation. The thickness of the steel poles of the dipole magnet was optimized to alleviate saturation within the poles. A 2D map of the field perpendicular to the pole tips can be seen in the middle panel of Fig.~\ref{fig:mag_comp}.

\subsubsection{Magnetic Field Measurement}

\begin{figure}[!htb]{\centering
\includegraphics[width=0.47\textwidth]{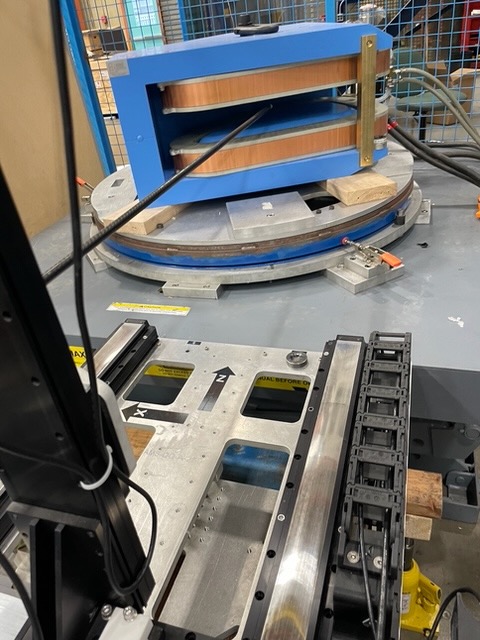}
\caption{\label{fig:Magtest}DarkLight dipole magnet being tested at TRIUMF.}}
\end{figure}

\begin{figure}
    \centering
    \includegraphics[width=\linewidth]{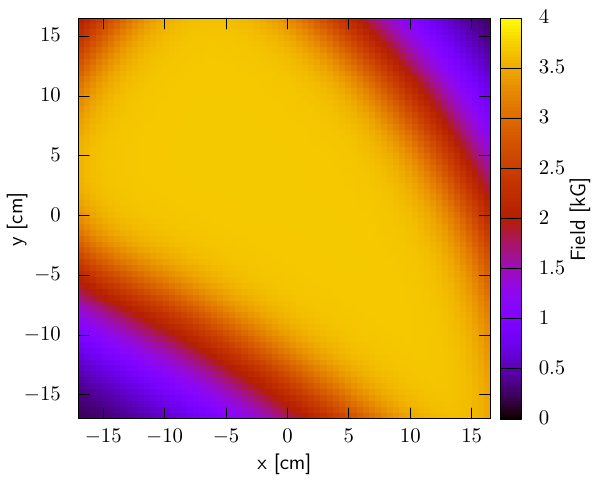}
    \includegraphics[width=\linewidth]{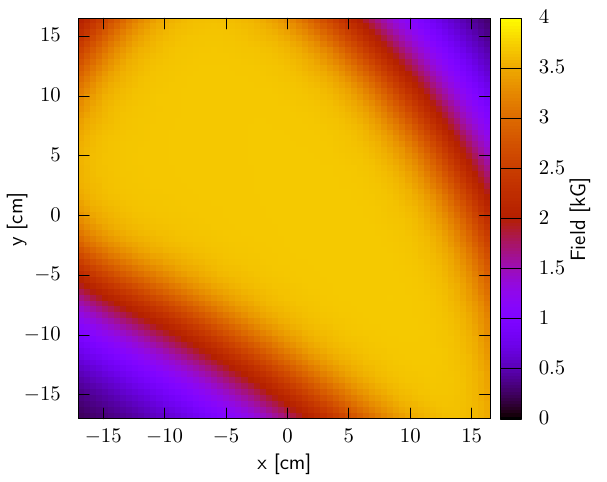}
    \includegraphics[width=\linewidth]{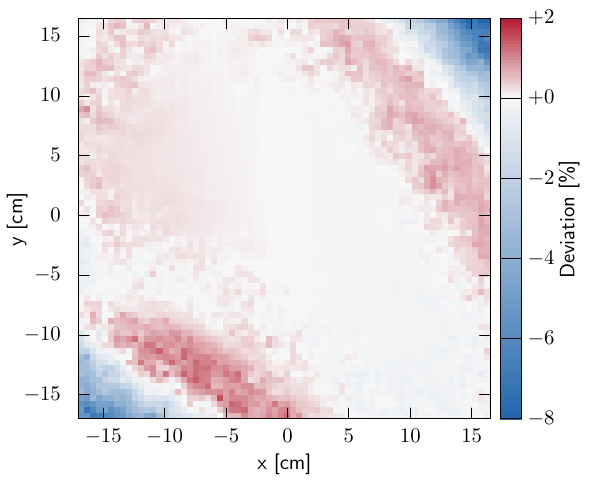}
    \caption{\label{fig:mag_comp} \textit{Top}: The measured field in Gauss. \textit{Middle}: The calculated field in Gauss from Ansys-Maxwell, with a scale factor of 1.063 applied. \textit{Bottom}: The deviation between the measured and scaled calculated fields. The range of the colour bar represents the maximum and minimum deviation.}
\end{figure}

The spectrometer dipoles were field mapped at TRIUMF to verify that the designed field is realised and to provide data for track reconstruction. A picture of a spectrometer being mapped is shown in Fig.~\ref{fig:Magtest}.

Results of the field mapping were consistent with the Ansys Maxwell calculations. A comparison of the measured and calculated fields can be seen in Fig.~\ref{fig:mag_comp}. The position of the calculated field was fit to the measured field to provide a precise comparison between the maps. It was found that the calculated field needs a uniform scale factor of 1.063 to best match the measured field. This scale factor is believed to arise from hysteresis effects in powering the magnet, as well as the limited knowledge of the orientation between the one-dimensional hall probe and the magnetic field that was measured. In the region of the pole tips, there is sub-percent deviation between the two fields. However, since the mapping was limited in the physical range accessible by the measurement station, and the mapping could only be performed in a single dimension, the Ansys Maxwell calculations were used in the Monte Carlo simulation and the analysis.

The field strength was monitored during the experiment using a Hall probe mounted between the pole tips on each magnet. In addition the current output of the power supplies was monitored. As the Hall probes are mounted to the magnet without radiation shielding, they are vulnerable to radiation damage over the course of the experiment. To calibrate the probes, we occasionally inserted a portable reference probe (FW Bell Model 5180) which is kept away from the radiation area.

\subsection{Dipole Vacuum Chambers}

\begin{figure}{\centering
\includegraphics[width=0.47\textwidth]{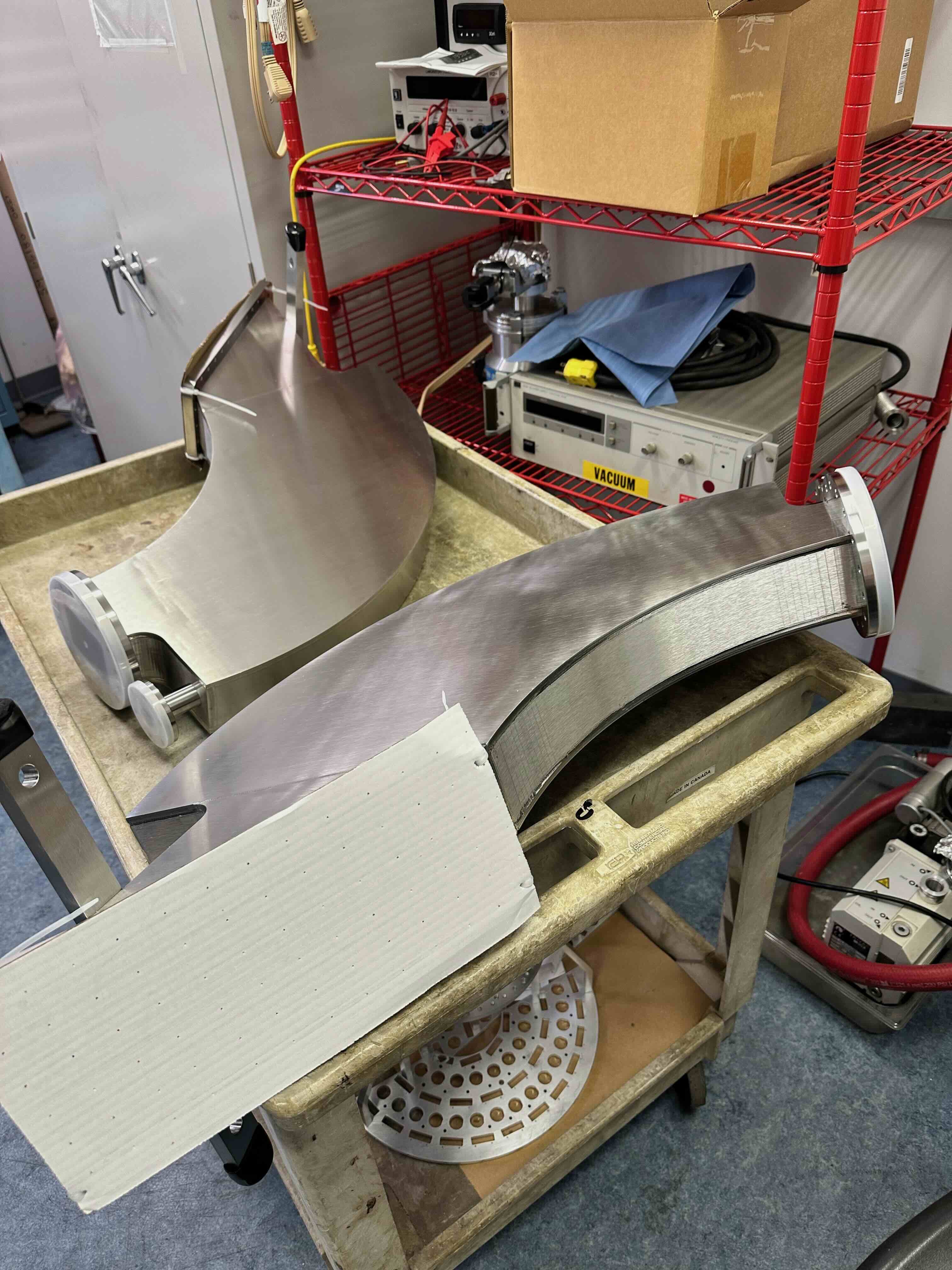}
\caption{\label{VacCham}Dipole vacuum chamber before installation into the dipole magnets.}}
\end{figure}

\begin{figure}{\centering
\includegraphics[width=0.47\textwidth]{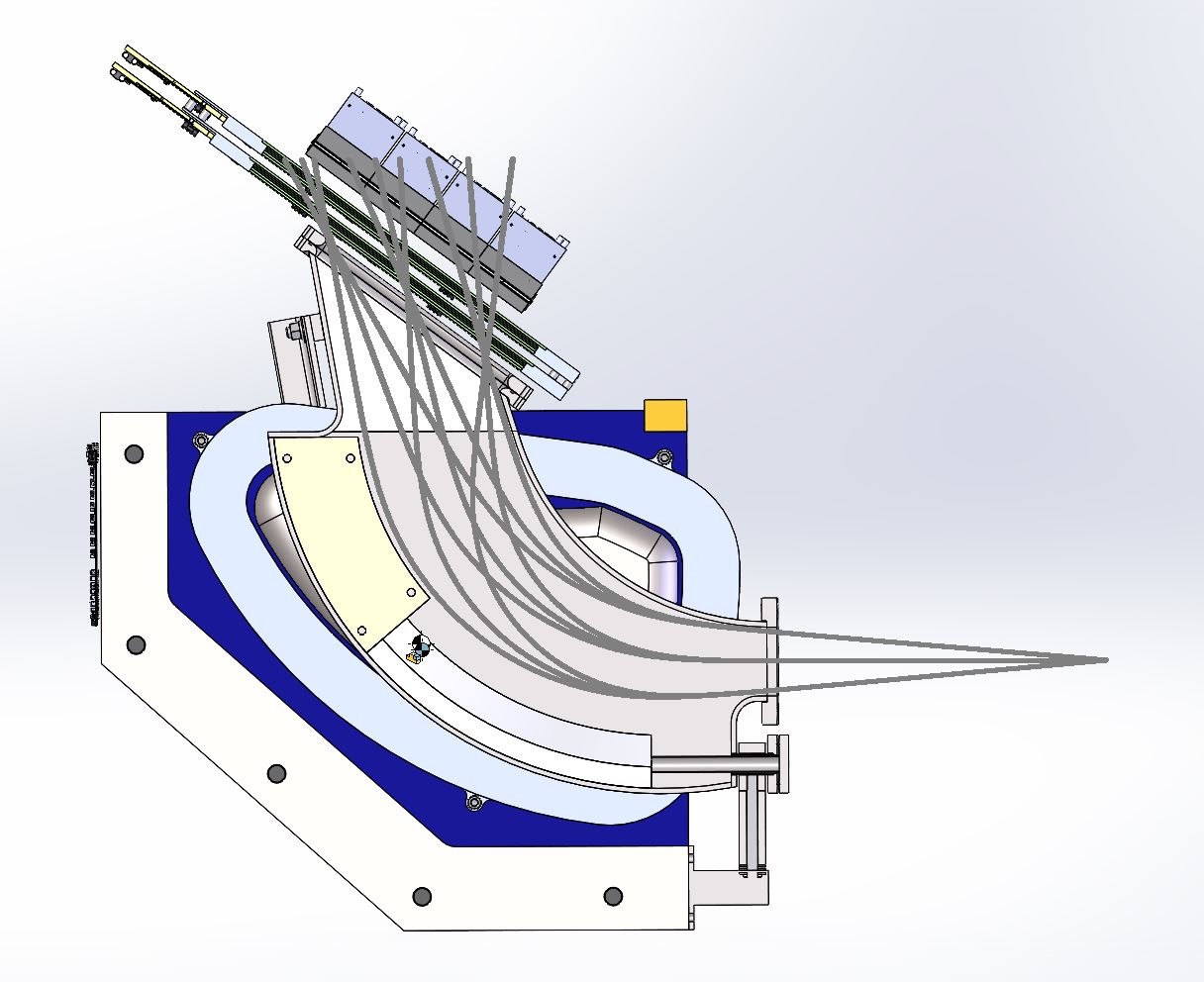}
\caption{\label{Vacuum}Cut away view of spectrometer showing the vacuum chamber with PEEK insert, the two GEM detectors and the trigger detector. Also shown are reference particle trajectories.}}
\end{figure}

A vacuum chamber for each dipole magnet was designed at MIT-Bates and manufactured by Ancorp.  Each chamber consists of a stainless steel vessel 78~mm in outside width (to fit between the 80~mm opening between the dipole pole tips).  The vacuum chambers walls are 
5~mm thick.  Each chamber couples to the scattering chamber vacuum flange via a flexible bellows. Additionally a short section of beamline that contains the tungsten collimators is inserted between the scattering and dipole chambers; further details are described in Sec.~\ref{sec:magnet_spectrometers}.  The two vacuum chambers are shown in~\cref{VacCham}.

The exit at the top of the vacuum chambers matches the angle of the focal plane and the vacuum is terminated by a Viton O-ring paired with a \SI{250}{\micro\metre} thick 5052 H0 aluminum window. This was pressure tested at MIT-Bates to four times atmospheric pressure.

The tungsten collimator restricts particles from the target from striking the sides of the vacuum chamber. The magnetic field bends the particles of interest through the exit window and onto the focal plane, and then through the two GEM and trigger detectors as shown in \cref{Vacuum}.

Not all particles entering the vacuum chamber are at the momenta set by the magnetic field.
In general the momenta of interest ({\it e.g.} for the resonant decay of the $A^\prime$ leptons) are significantly lower than the beam momentum. Thus elastically scattered particles will enter the vacuum chamber and will not be safely directed through the exit windows. Such particles will strike the back wall of the vacuum chamber and generate background in the detectors.  To minimize this effect the back wall of the vacuum chamber was extended and filled with 10~cm of PEEK (polyetheretherketone).  This material is compatible with high vacuum, radiation resistant, mechanically stable and has sufficient density to significantly absorb the higher momenta particles without creating large background in the detectors. 

The similar problem in the positron arm is not so severe as the elastic scattering particles are bent in the opposite direction. Nevertheless, a similar solution is employed for those cases when the positron arm is operated in electron polarity.

\subsection{Triple GEM Detectors}

\begin{figure}{\centering
\includegraphics[width=0.4\textwidth]{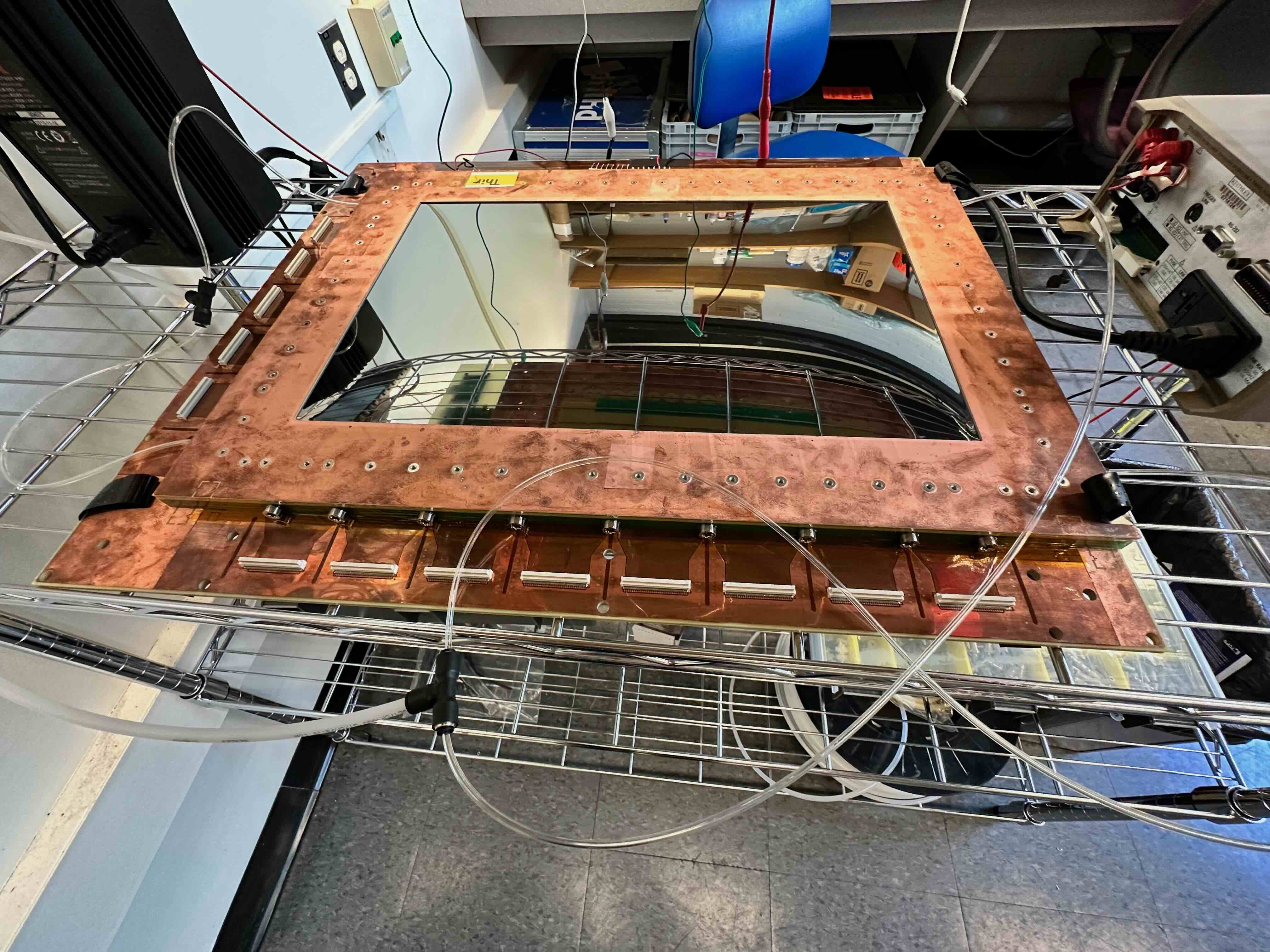}
\caption{\label{GEM}A single triple GEM detector at a test station before installation into the DarkLight experiment.}}
\end{figure}

\begin{figure}{\centering
\includegraphics[width=0.47\textwidth]{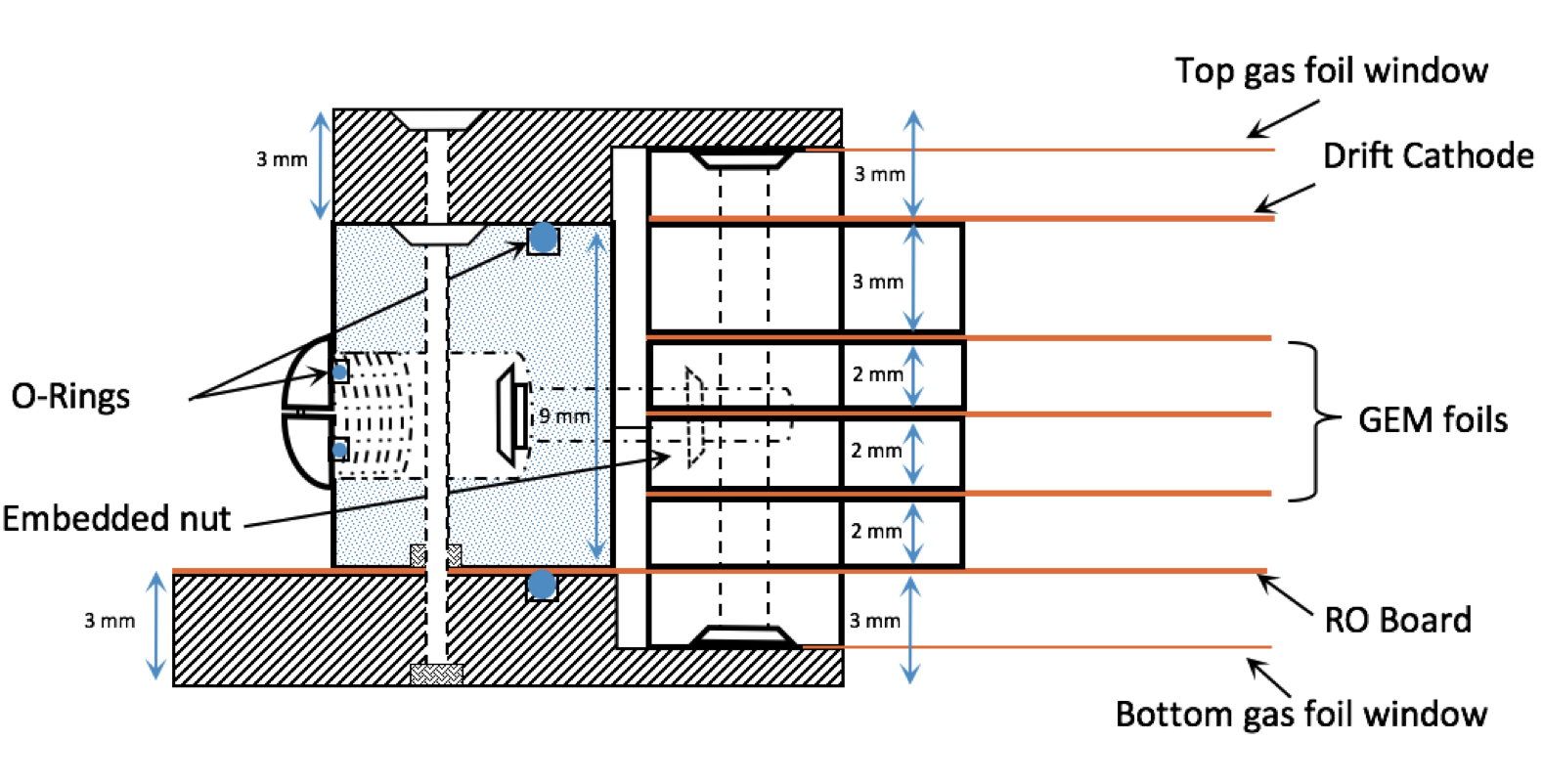}
\caption{\label{NS2}Cross section view of a triple GEM detector.}}
\end{figure}

Each spectrometer was instrumented with two layers of triple GEM detectors.  The first GEM detector on each spectrometer was situated close to the spectrometer focal plane and the second 20 mm above to allow for the determination of scattering angles.  The GEMs are read out via standard APV electronics. They were provided by the Hampton University group.  The APV readout is triggered by the scintillating strip hodoscope described in Sec.~\ref{sec:trigger_hodoscopes}.

\cref{GEM} shows one of the triple GEM detectors at a test station before installation into DarkLight.
The active area of each GEM is 25~cm in the non-dispersive direction and 40~cm in the dispersive direction. 
The APV boards and backplanes are not mounted in this view. A cross sectional view of a triple GEM detector is provided in \cref{NS2}, illustrating how the tension of the foil stack is achieved mechanically with an adjustable screw system outer frame pulling the inner stack outwards, adapted from CERN~\cite{ABBANEO201967}.

There are pressure volume foils at the entrance and exit of the GEM detector. These bulge due to the slight over pressure of the GEM gas in order to prevent the other foils from experiencing any pressure, and to prevent atmospheric gases leaking into the chamber. The 3~mm drift volume is where the track ionization occurs followed by three layers of GEM foils that amplify the electron showers by roughly $10^4$ before the cascade is collected on the readout layer.  The readout layer consists of a double sided pattern of $x$ and $y$ strips with \SI{400}{\micro\metre} pitch. The charge collected by each strip is read out by APV cards connected to a Multi-Purpose Digitizer (MPD) located in a VME crate.

The GEMs were produced by the Hampton University group for the original conception of DarkLight, situated at Jefferson Lab~\cite{PhysRevLett.111.164801,LEE201946}. 

\subsection{Trigger Hodoscopes}\label{sec:trigger_hodoscopes}

\begin{figure}
{\includegraphics[width=0.47\textwidth]
{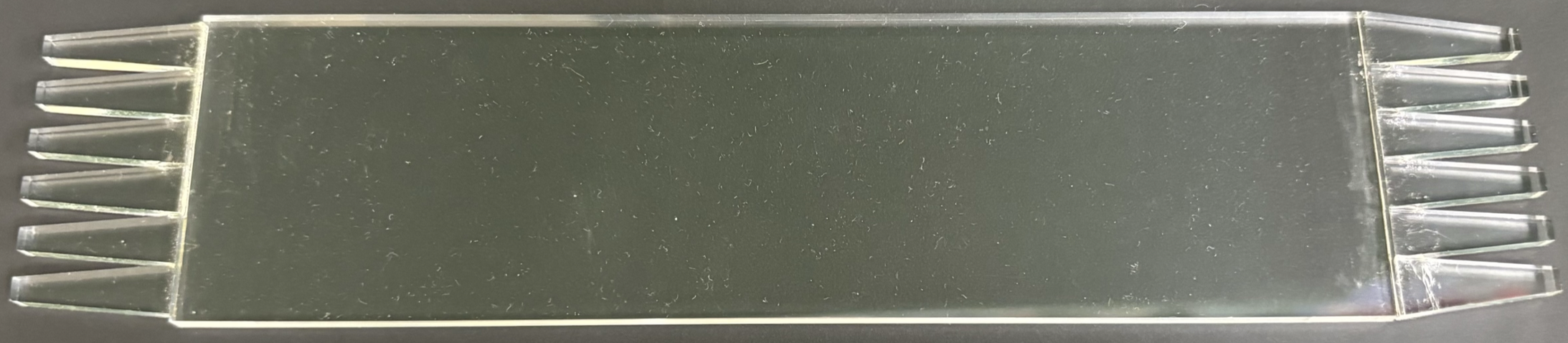}
\caption{\label{trigpaddle}An individual trigger bar with the light guides on either end. Figure from~\cite{Gelinas2025}.}}
\end{figure}

\begin{figure}
{\includegraphics[width=0.47\textwidth]
{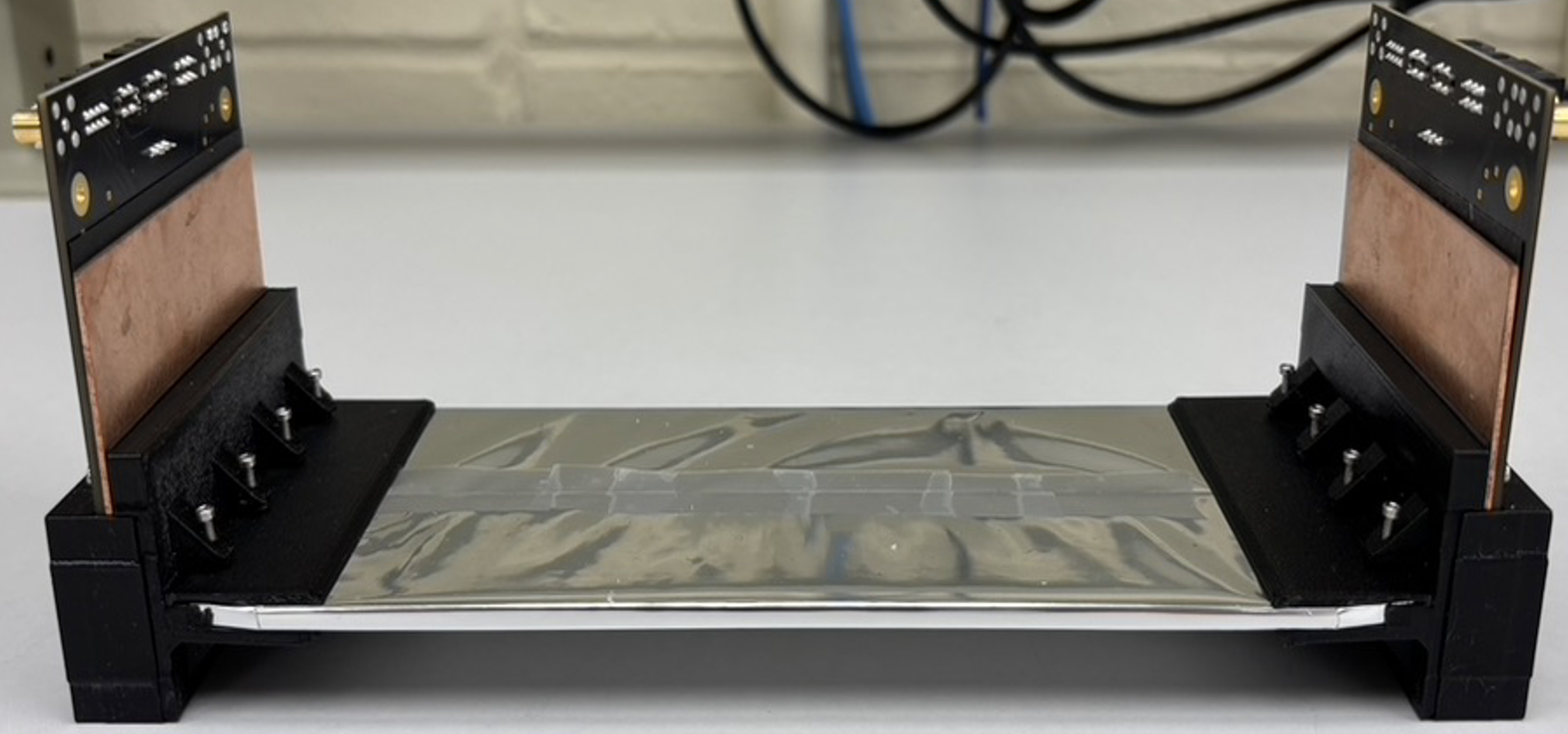}
\caption{\label{trigassembly} Two trigger bars connected with their plastic holder with SiPMs, readout electronics, and cooling plate to form a trigger paddle. Figure from~\cite{Gelinas2025}.}}
\end{figure}

\begin{figure}
{\includegraphics[width=0.47\textwidth]
{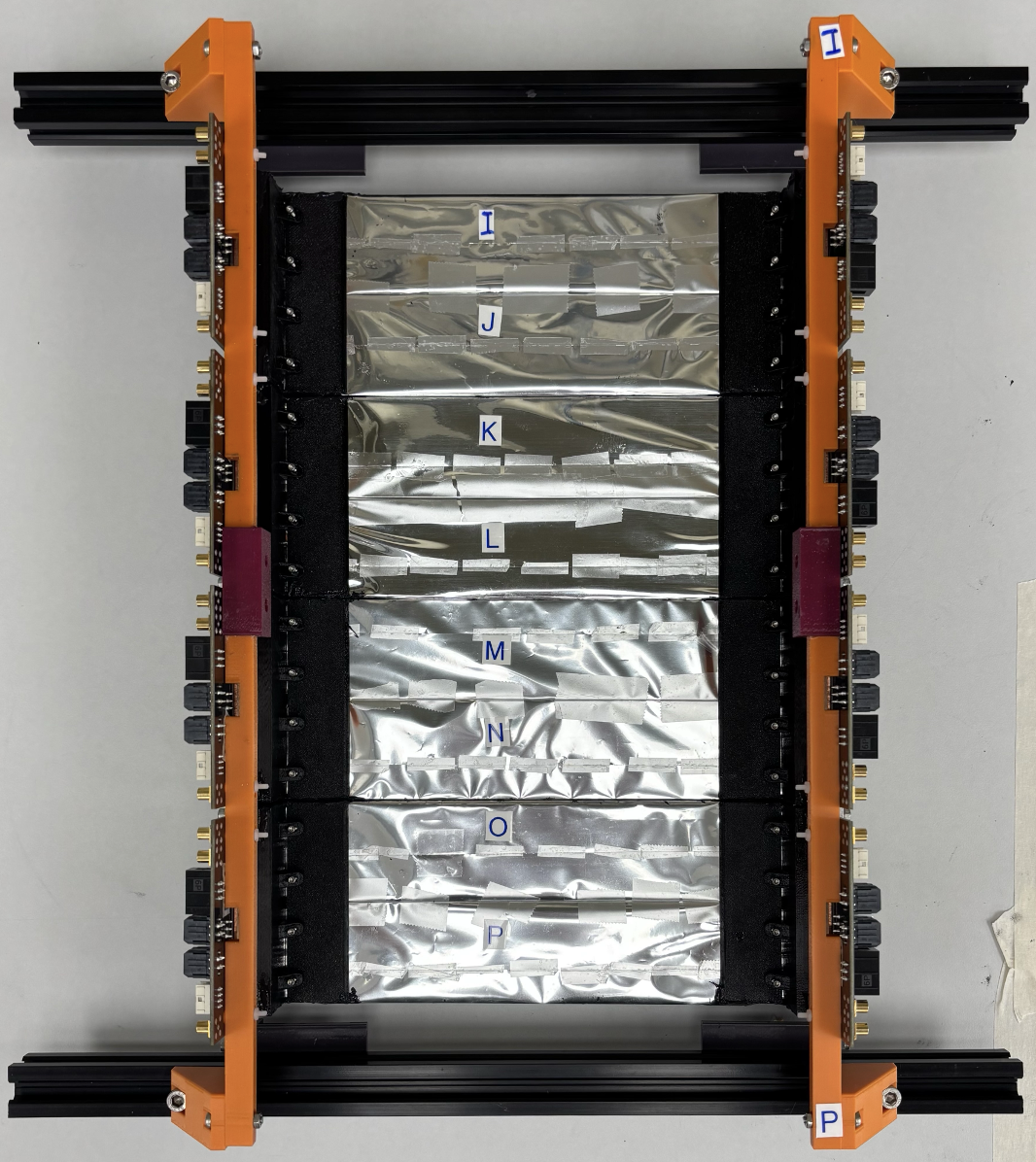}
\caption{\label{trigtop}Top view of a complete trigger hodoscope with four paddles. Figure from~\cite{Gelinas2025}.}}
\end{figure}

The APV GEM readout requires a trigger signal.  To provide this we have a pair of scintillating strip hodoscopes with SiPM readout on top of each spectrometer.  Signals from the left or right hodoscope, or a left-right coincidence, can be used to trigger the GEM readout.  

Each hodoscope consists of four individual scintillator paddles, each containing two individual scintillator bars, made from BC-404 plastic with a size of 170$\times$40$\times$3.2~mm$^3$. Each bar (see~\cref{trigpaddle}) has six tapered light guides attached at either end that are then coupled to SiPMs (3x3~mm$^2$  Hamamatsu S13360-3075PE). Two bars are combined in a support frame into a paddle that holds the SiPMs, readout electronics, and a copper cooling plate (see~\cref{trigassembly}).

To reduce accidental coincidences in the trigger logic, it is important to resolve the beam bunch clock of 650~MHz, at least at the analysis level. This timing information must be provided by the trigger detector, but can be corrected by the particle path length reconstructed from the tracking detector information. However, to reduce readout dead-time, it is important to be close to the ideal timing during data-taking. In order to facilitate this, the trigger channels can be timed individually, as the main time dispersion is generated by the momentum-dependent dispersion inside the spectrometers. 

\subsubsection{Trigger Hodoscope Electronics}

The six signals from each end of the trigger hodoscope counters were preamplified and summed before being amplified and discriminated (both leading-edge and trailing-edge) by an  FPGA-based TDC. The firmware of the FPGA allows us to generate triggers from $e^+e^-$ coincidences in very small coincidence windows, typically not possible with off-the-shelf coincidence units. The time-over-threshold was used to make time-walk corrections to improve the overall time resolution. The small coincidence time windows were required to identify coincidences on the level of individual bunches, thereby minimizing the recorded background. By calibrating timing offsets for each channel, the FPGA also allows us to easily adjust the timing of individual trigger bars, since the particle path length through the spectrometer, and thus, the time offset from the vertex, depends on the momentum. 

\subsection{Detector Shielding} \label{sec:detector_shielding}

In addition to the shielding issues presented in the section on the beam interaction with the target (Secs.~\ref{sec:BeamTarget} and ~\ref{sec:target_shielding}) there is a significant background electromagnetic radiation rate generated by the RF cavities, even when the beam is off.  Field emission in the cavities produces low energy electrons which can be accelerated by the RF cavities.  These then strike beamline elements and produce high background rates in the electron hall.  Recent tests showed rates on the order of several hundreds of kHz even when the detectors were directly screened from the RF cavities by concrete shielding blocks and lead sheets.  The rate was proportional to the cavity high voltage and consists mostly of low energy X-rays.

To address the radiation that would strike the detectors both from beam-target interactions and from production in the RF cavities, a frame was constructed to support 1~cm of lead around the GEMs and trigger hodoscope. The full shielding implementation can be seen in the bottom panel of Fig.~\ref{fig:dl_fullinstall}
\subsection{Installation}

\begin{figure}
    \centering
    \includegraphics[width=\linewidth]{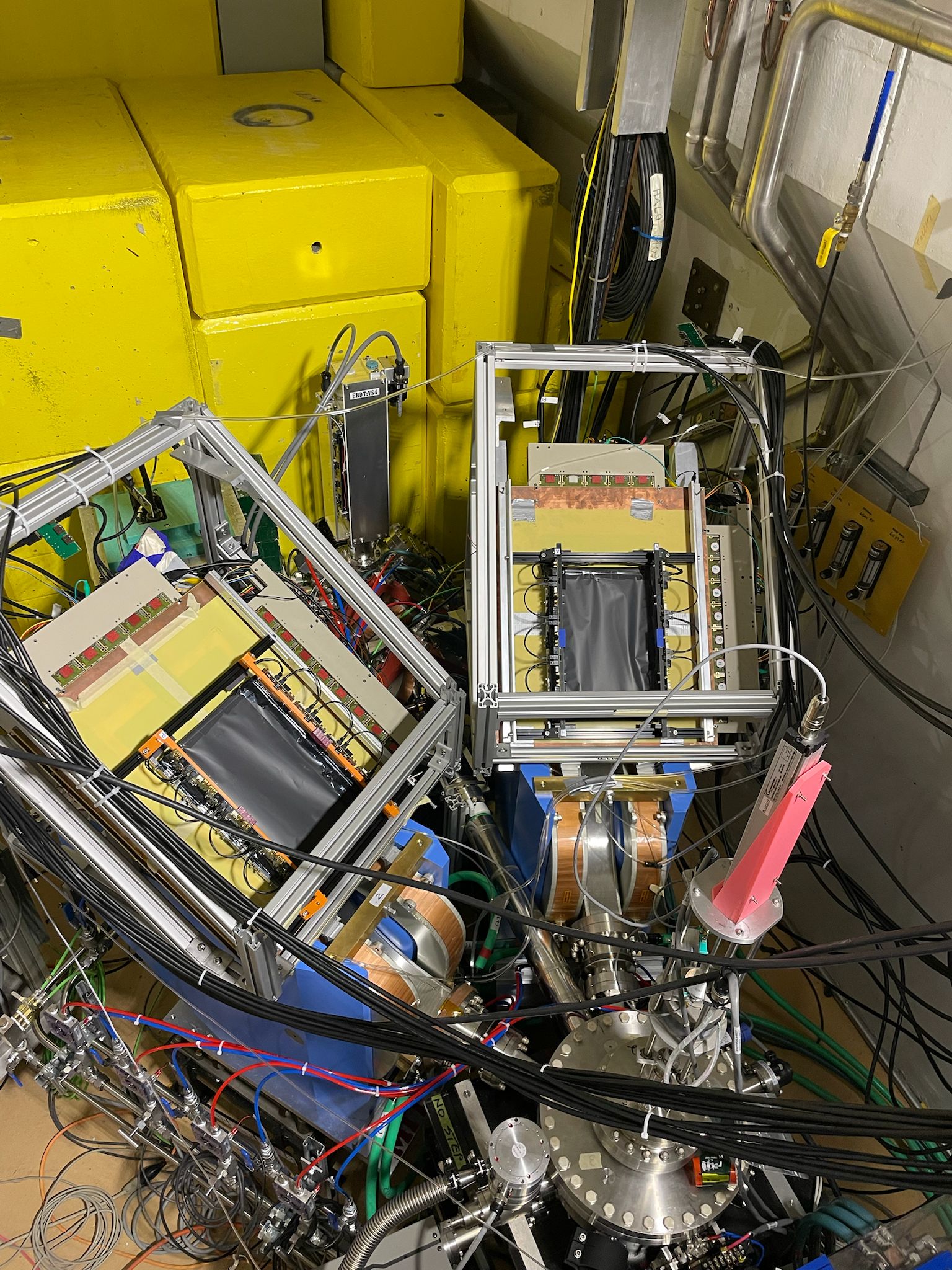}\\
    \includegraphics[width=\linewidth]{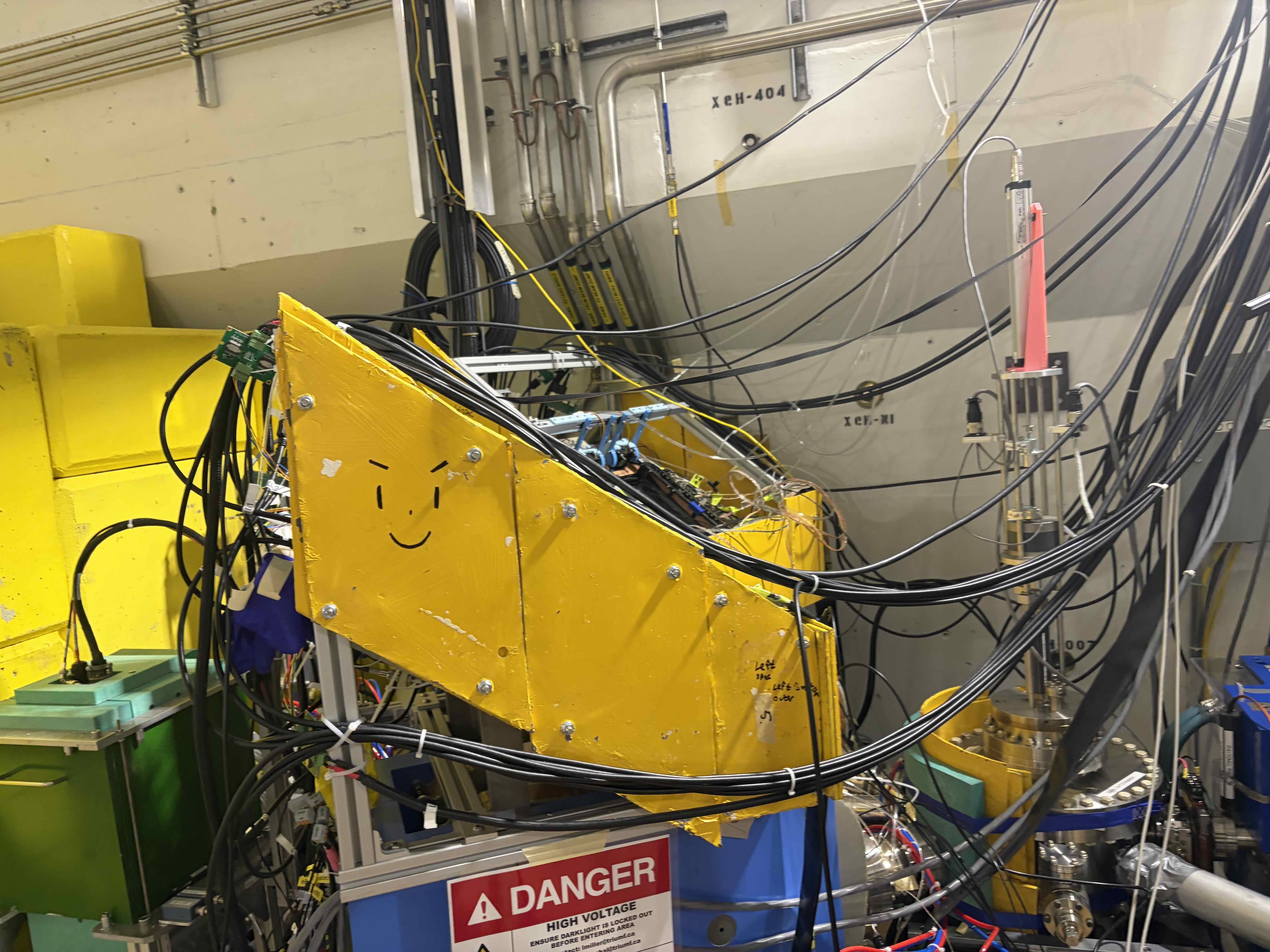}
    \caption{Two photographs of the installed DarkLight experiment. \textit{Top}: The trigger scintillators can be seen in black within the detector shielding frame and on top of the GEMs. The blue dipoles can be seen below the detectors. The large yellow blocks in the background are the concrete shielding blocks from the beam dump. The tall object in the front right of the figure is the target ladder assembly, including the driver motor and support structures. At the bottom of the figure is the target chamber. The exit beamline can be seen between the two spectrometers. \textit{Bottom}: A side view of the installed experiment. The yellow sheets are lead shielding. In the bottom right of the figure, the light green sheets around the target chamber are borated polyethylene and sit in front of the target lead shielding.}
    \label{fig:dl_fullinstall}
\end{figure}

The DarkLight experiment was installed in the TRIUMF e-linac in the summer of 2025. Photographs of the completed installation can be seen in Fig.~\ref{fig:dl_fullinstall}.

\section{Data Acquisition and Slow Control}

The GEMs are read out via APV25 ASICs connected to MPD4 digitizer boards, one board per GEM. The trigger detectors are read out via a custom version of the Chronobox, a Cyclone-V FPGA system that TRIUMF has developed for the CERN Alpha-g collaboration~\cite{cernalphag}. 
 
The Chronobox is designed to be a streaming system. To accommodate the triggered nature of the APV25, the system also generated triggers with optional pre-scaling and implements a busy system that suppresses the generation of triggers during the MPD4 readout.

For synchronization, the trigger system provides a 40 MHz clock to all MPD4 boards. This clock drives a free running counter that is sampled and added to the data stream at every trigger.

The readout of the MPD4 modules is performed via VME64 block reads using a VME CPU. Using a custom MPD4 firmware, the data processing inside the MPD4 is essentially dead-time free, and maximum event rates are a function of data transport speed and event size. We achieve close to the maximum line rate of the 1 GBit/s network interface of the VME CPU.

The DAQ software system is MIDAS based, but replaces the default file writer with a custom system that writes COOKER-compatible ROOT files~\cite{ROOT_NIMA_1997} (see below) and interfaces with the slow control system. 

The slow control system is based on EPICS~\cite{epics}, using the framework developed for the OLYMPUS \cite{MILNER20141} experiment. A EPICS Soft-IOC communicates with the various hardware devices. Various support modules import data from the accelerator control system and the DAQ system into the IOC, which re-exports all channels as records with a common naming scheme. The IOC also implements basic slow control logic. Record changes are automatically pushed to a PostGreSQL database that serves as a historical archive. The DAQ also writes the complete slow control data stream to the ROOT file.

Control GUIs are implemented via a web interface. A python app using the FLASK package implements a web API that provides user account management, archive access, a pull-based record update process, and a way to affect record changes for authorized users. A set of templated web sites display the information, provides push-button control elements etc.

\section{Simulation and Analysis}
\label{sec:Simulation}
For simulation and event reconstruction studies, the DarkLight experiment uses the COOKER framework, which has previously been utilized in the OLYMPUS~\cite{MILNER20141}, TREK~\cite{Dongwi_2022}, TPEX~\cite{alarcon2023twophotonexchangetpex}, and MUSE~\cite{gilman2017technicaldesignreportpaul} experiments. Developed in C++ and based on the ROOT data analysis framework, COOKER provides a modular software architecture that supports a wide range of analysis plugins. It covers the full analysis chain, including event generation, Geant4~\cite{AGOSTINELLI2003250} detector simulation, event reconstruction, and data analysis within a unified software framework.

Within the COOKER framework, multiple event generators are used to simulate different processes and produce high-statistics samples for detector simulations and physics projection studies.
The Mainz generator~\cite{Beranek:2013yqa} is used to calculate the cross section for $ep \rightarrow ep\gamma^\ast \rightarrow epl^+l^-$ at leading order with all leading order radiative corrections. The production of an $A^\prime$ signal decaying to a lepton pair is kinematically indistinguishable from the production of a $\gamma^\ast$ with the same effective mass. The Mainz generator can therefore be used to generate both signal samples and irreducible background samples. For signal samples, $\gamma^\ast$ events are generated in a very narrow mass window around the $A^\prime$ mass of interest, and only for leading order diagrams wherein a virtual photon directly decays to an $l^+l^-$ pair. The cross section is normalised to that of a dark photon by a scale factor dependent on the kinetic mixing $\epsilon$. For background estimation, the same generation takes place, but all leading-order diagrams are included, the $\gamma^\ast$ mass is not restricted, and the cross section needs no rescaling. The use of $\gamma^\ast$ production as a proxy for dark-photon production is an established method~\cite{Ilten:2016tkc}.
The generator developed for OLYMPUS~\cite{OLYMPUS:2016gso} is used to generate electrons which, after scattering in the target, radiate a hard photon and lose sufficient energy to enter the electron spectrometer acceptance. As with the Mainz generator, the outgoing particles are placed within the detector acceptance, and the cross sections for the relevant event is then computed. The radiative cross section calculation uses the Bethe-Heitler approximation for photon radiation from the electron and the Born approximation for photon radiation from the proton, and includes additional tail corrections.
For experiment commissioning studies, a M\o{}ller/Bhabha event generator~\cite{Epstein:2016lpm} and a radiative elastic electron-carbon scattering generator~\cite{eCgen} are used.

The DarkLight detector setup is fully simulated in Geant4 using a magnetic field map calculated with ANSYS Maxwell. Detector responses to particle interactions are further modeled in COOKER through dedicated digitization plugins that reproduce the expected detector performance. The GEM simulation utilizes Garfield++~\cite{garfield} and HEED~\cite{heed} to model charge transport and signal spreading across multiple readout channels, particularly for tracks traversing the detector at large angles. Within the COOKER analysis framework, detector positions and hit coordinates in various experimental coordinate systems can be obtained directly from the simulation. Because the simulation closely reproduces the detector geometry, magnetic field, and detector response of the experiment, it can be used for detailed comparisons with experimental data. It also provides training and validation samples for the development of reconstruction algorithms, including vertex reconstruction and event reconstruction methods.

DarkLight adapts multiple methods to reconstruct the reaction vertices and particle kinematics from detector hit information after the dipole modifies the particle trajectories. As the baseline approach, a traditional polynomial fitting method of the form
\begin{equation}
f_i=\sum_{a,b,c,d} \alpha_{a,b,c,d} x^a y^b dx^c dy^d,
\end{equation}
is used to approximate the dipole transfer matrix. This parametrization maps the lower GEM hit positions, $x$ and $y$, together with the corresponding position differences between the two GEMs in a spectrometer arm, $dx$ and $dy$, all expressed in the detector local coordinate system, to the electron and positron kinematic variables, $f_i$, at the reaction vertex. For each kinematic variable, a polynomial basis up to a specified maximum order, $a-d$, is constructed. The coefficients, $\alpha_{a-d}$, are determined independently by fitting to the corresponding transfer matrix, and polynomial terms that have a negligible impact on the reconstruction performance are removed to optimize the parametrization. To explore and compare alternative reconstruction techniques, machine learning approaches are also investigated, including gradient boosting decision trees and deep neural networks.

\section{Expected Count Rates}

\begin{figure*}[htb!]
\centering
    \includegraphics[width=\linewidth]{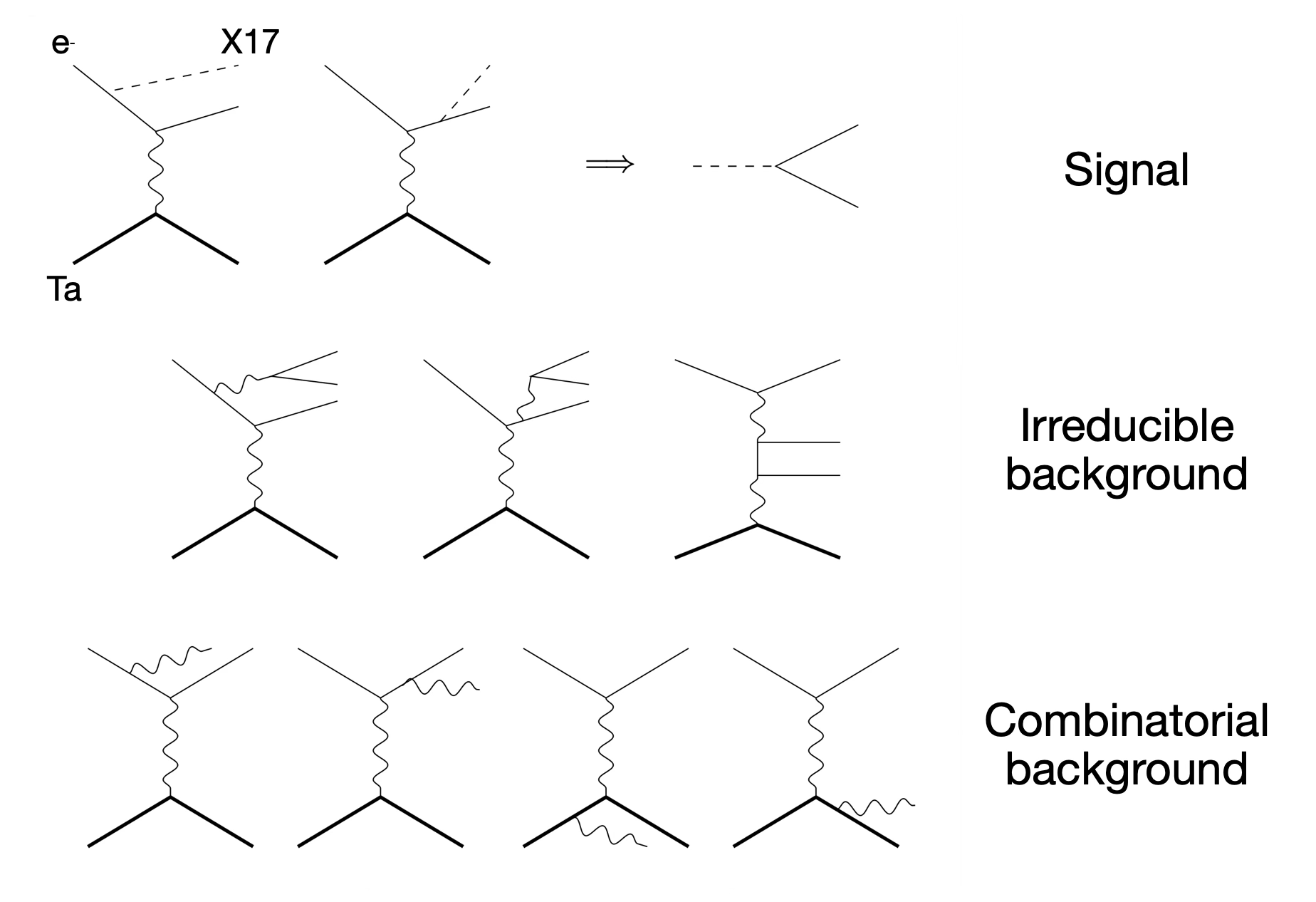}
\caption{Feynman graphs for the signal and dominant background processes. First row: The $A^\prime$ is produced off the incoming or outgoing lepton and then decays into an $e^+e^-$ pair. Second row: The irreducible QED background processes produce an $e^+e^-$ pair via an intermediate virtual photon. Third row: The trigger condition can be fulfilled by accidental coincidences of an electron from radiative elastic scattering combined with a positron of the irreducible QED background. For the proposed kinematics and luminosity, this is the dominant background process.}
    \label{figfeyn}
\end{figure*}

\begin{table}
    \centering
    \caption{Background rates for the proposed measurement settings. For both the settings a beam current of \SI{150}{\micro\ampere} is assumed. The 13@30 setup has the electron (positron) spectrometer arm at an angle of $36^\circ$ ($20^\circ$) from the beam axis, while the 17@50 setup has both spectrometer arms at an angle of $21^\circ$ from the beam axis.}

    \begin{tabular}{c|c |c |c|c}
        Setup&Irr.& Singles& Singles& Random\\
        & QED& $e^+$&$e^-$&coinc.\\
        \hline
        \hline
        13@30 & 3.1 Hz & 18.6 kHz & 5.4 MHz & 153 Hz \\
        17@50 & 16.2 Hz & 15.4 kHz & 19.6 MHz & 464 Hz
    \end{tabular}
    \label{table:rates}
\end{table}

To first order, the signal and irreducible QED background processes are described by the Feynman diagrams depicted in Fig.~\ref{figfeyn}.
We estimate count rates for the $A^\prime$ signal and the irreducible QED background using the Mainz generator as described in Sec.~\ref{sec:Simulation}. 
In addition to the QED background, the high luminosity leads to additional background from random coincidences, as demonstrated in the third row of Feynman diagrams in Fig.~\ref{figfeyn}. 
In this case, the positron spectrometer detects a positron from the irreducible QED background, however the corresponding electron has kinematics that are not detected by the electron spectrometer -- the expected rate for this to occur is significantly higher and typically in the tens of kHz.
The coincidence condition is then fulfilled with a second scattering reaction in the same time window, producing an electron in the electron spectrometer acceptance. The dominant process here is elastic scattering with initial- or final-state radiation. The corresponding rates were estimated via the OLYMPUS generator~\cite{OLYMPUS:2016gso}. The magnitude of this random coincidence background depends on the effective coincidence window. For coincidences times longer than the bunch spacing, the time window is given by the time resolution of the spectrometers. However, when individual bunches are resolved a further increase in time resolution has no benefit, as all scattering events from one bunch essentially happen at the same time -- the bunch length is too short to resolve below that. For the rate estimate, we assume that we can resolve the 650~MHz bunch frequency, requiring timing resolution of $\mathcal{O}$(0.5~ns).

Table \ref{table:rates} gives the expected background rates for the setup at 30~MeV beam momentum, aiming at a $A^\prime$ mass of 13~MeV (13@30), as well as a future setup with a 50~MeV beam energy targeting a $A'$ with a mass of 17~MeV.  As can be seen, for the proposed kinematics and beam conditions, the random coincidence background dominates. It is important to note that this background scales with the luminosity squared ($\mathcal{L}^2$). The figure of merit (FOM) is given by the number of signal events divided by the square root of the background events. Thus, for luminosities in which the accidental coincidence background dominates, the FOM is independent of $\mathcal{L}$ and only scales with the measurement time. The reach cannot be improved with a further increase in instantaneous luminosity.

\begin{table}
    \centering
    \caption{Elastic scattering count rates into each spectrometer with Carbon and Tantalum targets assuming a \SI{1}{\micro\ampere} beam current on a \SI{1}{\micro\metre} target foil.}
    \label{tab:elasticrates}
    \begin{tabular}{c|c|c|c|c}
        Energy&\multicolumn{2}{c|}
        {Rate on C [kHz]}
        &\multicolumn{2}{c}
        {Rate on Ta [MHz]}\\
        {[MeV]}&$20\degree$&$36\degree$& $20\degree$&$36\degree$\\
        \hline
        10 & 369 & 66.6 & 28.8 & 4.63\\
        15 & 168 & 32.7 & 12.8 & 2.07\\
        20 & 97.4 & 20.5 & 7.22 & 1.17\\
        25 & 64.1 & 14.9 & 4.63 & 0.75\\
        30 & 45.9 & 11.9 & 3.22 & 0.53\\
    \end{tabular}
\end{table}

The elastic count rate on Tantalum in the range of the available electron beam energies of 10-30 MeV, with \SI{1}{\micro\ampere} current, and for a 2.813 (4.834) msr solid angle acceptance spectrometer at 20$^{\circ}$ (36$^{\circ}$), is between $\approx$ 0.5-28 MHz, see Table~\ref{tab:elasticrates}. This rate is beyond the timing resolution of the GEM detectors. The magnetic field will instead be probed with elastic scattering on Carbon, which, for similar assumptions on the beam current, energy, and acceptance, ranges between $\approx$ 11 kHz - 370 kHz.

\section{Spectrometer Angle Dependence}

\begin{figure}{\centering
    \includegraphics[width=\linewidth]{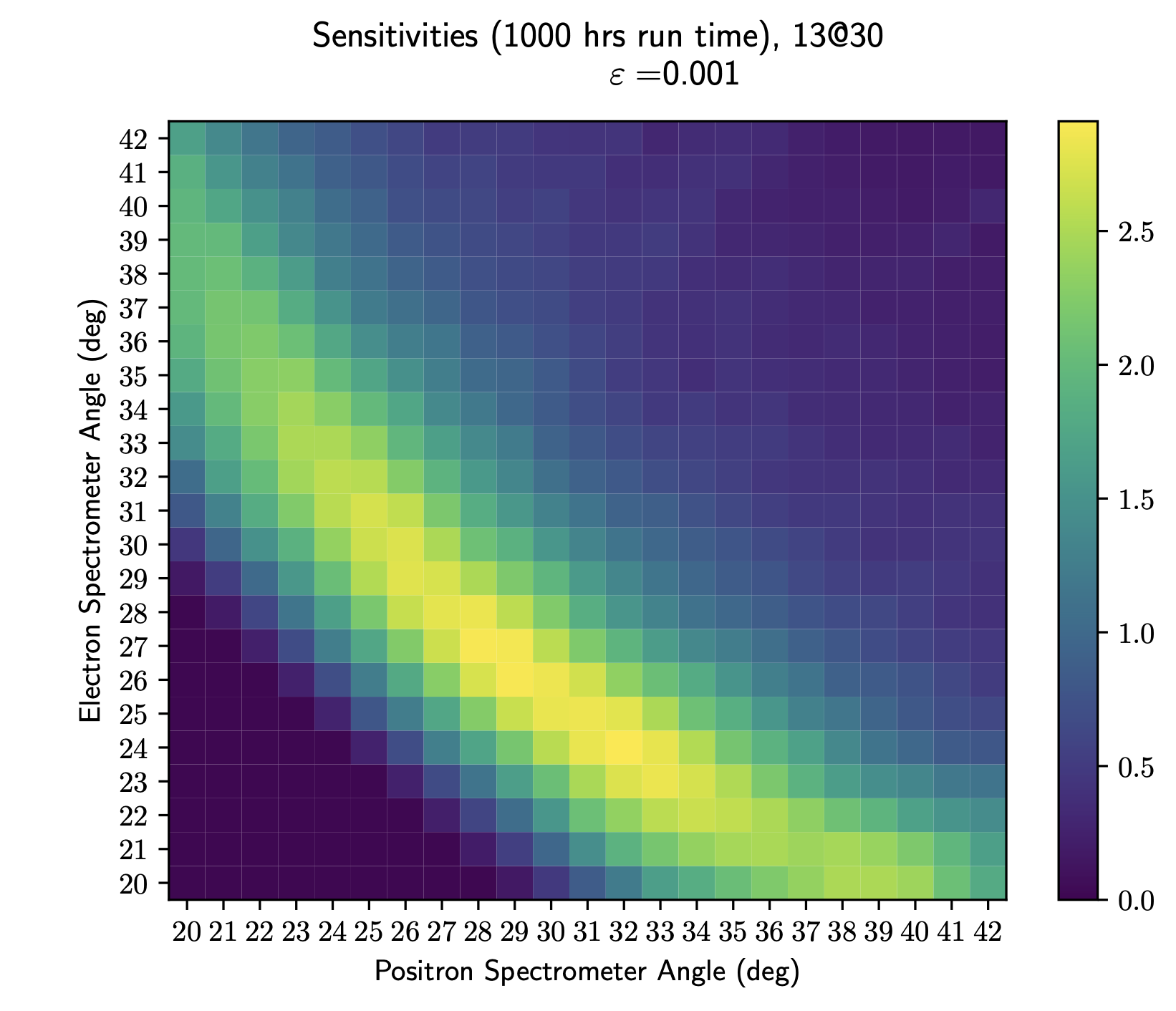}\\
    \includegraphics[width=\linewidth]{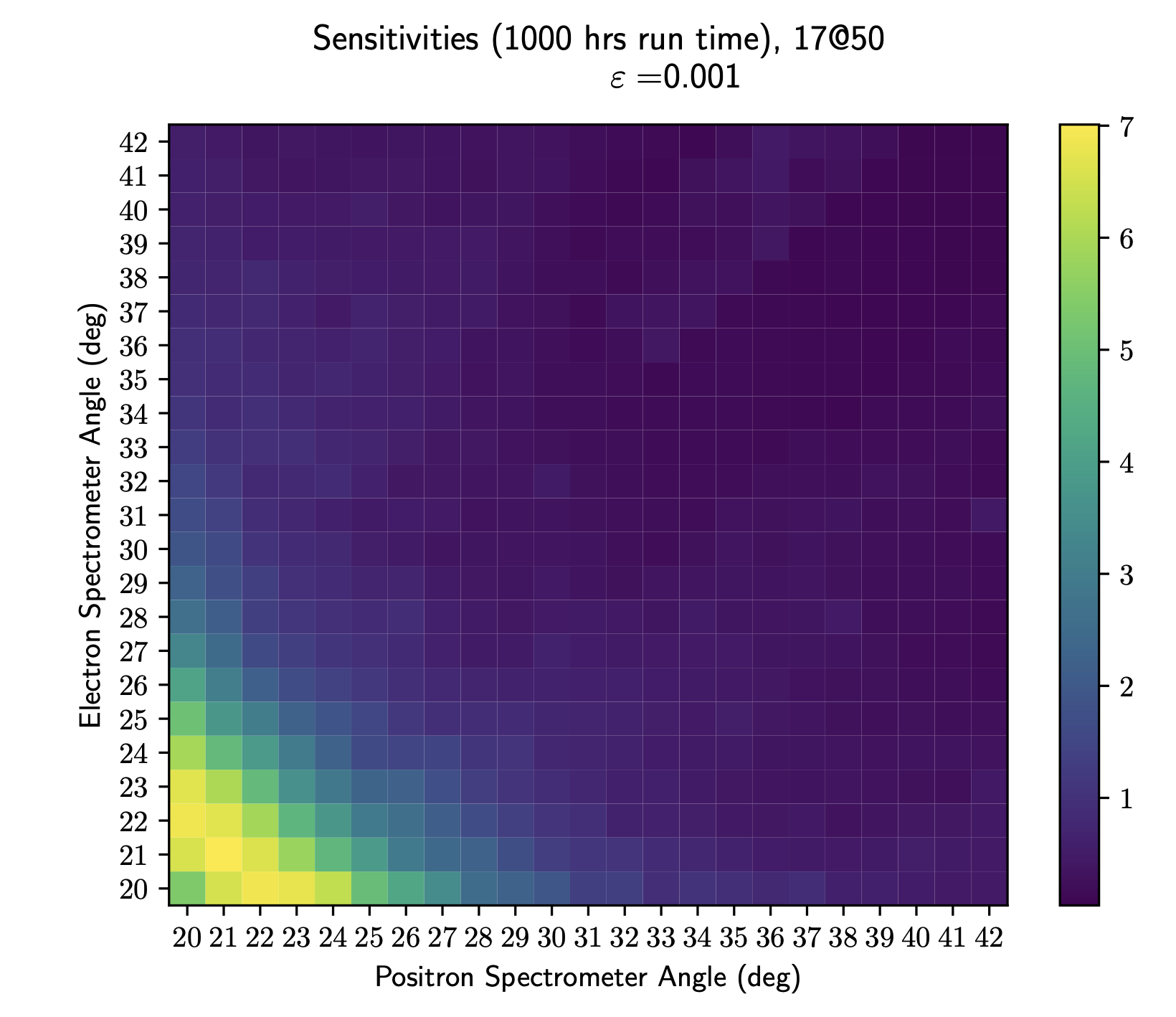}
    \caption{\label{fig:sens_angles}\textit{Top}: Projected sensitivity as a function of electron and positron spectrometer angles from the beamline for a 13 MeV $A^\prime$ with a beam energy of 30 MeV. \textit{Bottom}: Sensitivity to a 17 MeV $A^\prime$ as a function of spectrometer angle for 50 MeV beam energy.}}
\end{figure}

The count rates determined in the previous section can be used to estimate the sensitivity of the experiment to an $A^\prime$ as a function of the angles of the two spectrometer arms. For a predicted signal yield of $s$ and a background yield of $b$, considering only statistical uncertainty, the sensitivity is defined as~\cite{Cowan2011Asymptotic}
\begin{equation}
    \mathcal{S} = \sqrt{2((s+b)\ln(1 + s/b)-s)}.
\end{equation}
Using this expression, we calculate the projected sensitivity for each combination of electron and positron spectrometer angles. The top panel of Fig.~\ref{fig:sens_angles} shows the resulting sensitivity for a run time of 1000 hours at \SI{150}{\micro\ampere} beam current and 30~MeV beam energy with a target $A^\prime$ of mass 13 MeV and $\epsilon^2=10^{-6}$. Here $\epsilon^2$ represents the kinetic mixing parameter of the new boson. Maximum sensitivity for the 30 MeV experiment is reached with one spectrometer at approximately 29$^{\circ}$ and the other at approximately 27$^{\circ}$. However, due to space constraints in the experimental hall, the maximum angle for one spectrometer is 20$^{\circ}$. This leads to the other spectrometer being placed at approximately 36$^{\circ}$.
Changing either the beam energy or the target $A^\prime$ mass changes the kinematically favoured region. 
The bottom panel of Fig.~\ref{fig:sens_angles} shows the corresponding sensitivity for an $A^\prime$ of 17 MeV mass and a 50 MeV beam. For the future 50 MeV experiment, the 36$^{\circ}$ spectrometer will need to be moved closer to the beamline.
Although the ideal angle between the two spectrometers is roughly fixed by the kinematics of the $A^\prime$ decay, an orientation with the positron spectrometer near the beamline and the electron spectrometer farther away is favoured due to the relative reduction in random scattering backgrounds. 
\section{Projected Reach}

\begin{figure}
{\centering
\includegraphics[width=0.47\textwidth]{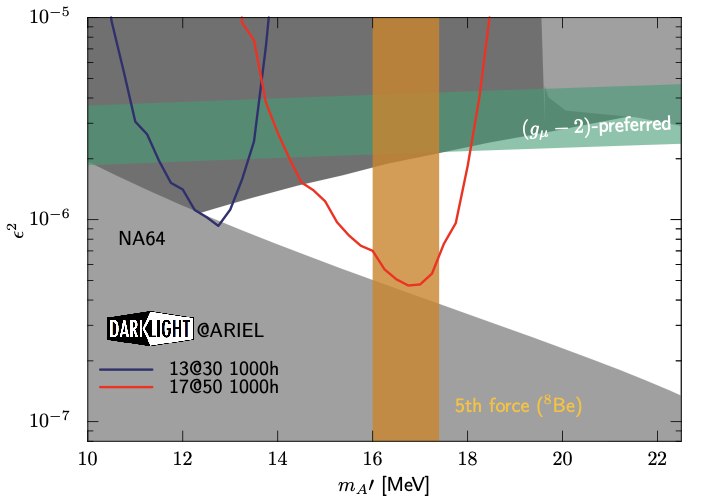} }
\caption{The projected reaches on a linear plot for three separate data taking runs: 13@30 (dark blue) $-$ a 1000~h run at 30~MeV optimized for $m_{A^\prime}$ = 13~MeV; 17@50 (dark red) $-$ a 1000~h run at 50~MeV optimized for $m_{A^\prime}$ = 17~MeV. Light gray areas are excluded by other experiments sensitive to a lepton coupling. The dark gray area is excluded by electron g-2 only.}
\label{reach1}
\end{figure}

For smaller $\epsilon^2$, the background and total signal are visually indistinguishable. However, the statistical uncertainty in the measured background becomes very small, and deviations are still detectable if the shape of the background below the peak is understood. Since random coincidences dominate the background, the pure random background will be accurately measured by the experiment itself by mixing electron and positron spectrometer data from different events. This mixing destroys all correlations between the spectrometers, generating a pure sample of the random coincidences. Since, in principle, every combination of events $i\neq j$ can be used, the  statistics grows quadratically with the recorded number of events. 

To estimate the experimental reach, we integrate the background over the expected signal width ($\pm 1.7 \sigma$), and determine the value of $\epsilon^2$ for which the signal would be bigger than a 2$\sigma$ fluctuation of the background. Figure \ref{reach1} shows the achievable reach for the two settings assuming 1000 h beam-time at each energy.

\section{Commissioning}\label{sec:commissioning}

\begin{figure*}
    \centering
    \includegraphics[width=\linewidth]{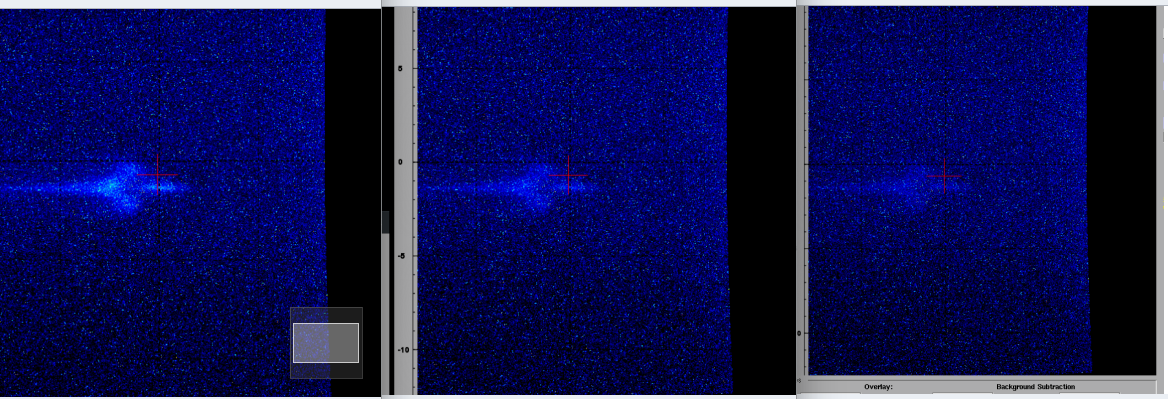}
    \caption{A picture of the beam spot at different cathode temperatures. The dark current can be seen as a horizontal line to the left and right of the figure. The left side of each image corresponds to the low-momentum side of the beam line. Left: Cathode set to 5.3 V. Middle: Cathode set to 5.2 V. Right: Cathode set to 5.0 V.\label{fig:viewscreen}}
\end{figure*}

An initial DarkLight commissioning run took place between November 2025 -- February 2026. There were two major data taking campaigns. The first took place from November to January where the beam energy was limited to a maximum of 10 MeV because the high-energy cryomodule responsible for energies above 10 MeV was inoperable. A replacement high-voltage control module was installed in mid-January which enabled the second data-taking campaign. This second campaign filled the remainder of the beam time and the beam energy could varied between 15~MeV -- 30~MeV. 

In both campaigns, the beam current could be varied between \SI{1}{\micro\ampere} -- \SI{100}{\micro\ampere} with a duty factory from 0.05~\% -- 2~\%. The overall beam power was kept below 100~W to prevent overheating in the beam dump. In principle the beam can be operated in a continuous wave mode, however for diagnostic and commissioning measurements the beam was operated in pulsed mode to provide a lower average current and enable a simpler analysis of data.

A future campaign will be needed to finish commissioning the experiment at the higher intensities useful for a BSM search.

The first priority of the measurement campaigns was to identify elastic scattering. As the energy of this process is known, it provides a useful calibration point to study the linearity of the magnet.

It was discovered that the beam had a significant off-momentum background. After investigation, it was determined that the cathode at the source of the electron gun had evaporated BaO coating onto the screen used to ``chop" and create the pulsed beam. This BaO coating was close enough to the cathode such that the radiated heat from the cathode caused the screen to emit background electrons, a ``dark current". These dark current electrons are emitted off-phase from the primary beam, and with an unknown and varying energy. These electrons were transmitted along the beam line and were transported toward the DarkLight target ladder. This background could be reduced, but not eliminated, by lowering the temperature of the cathode. A picture of a viewscreen upstream of the DarkLight target is shown in Fig.~\ref{fig:viewscreen}. In addition to lowering the cathode temperature, a gate was incorporated into the trigger system so that an event is only triggered during a time when an electron bunch passes through the target. The cathode will be replaced before the next commissioning run.

\subsection{Detector Performance}

\begin{figure}
    \centering
    \includegraphics[width=\linewidth]{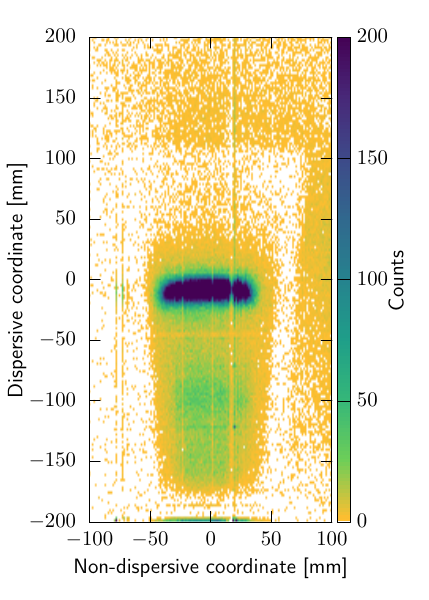}
    \caption{GEM hitmap showing the elastic line from 10 MeV electron scattering. The trapezoidal spectrometer exit flange can be seen as a deficit of clusters between -200 (-50) mm and 100 (50) mm in the (non-)dispersive coordinate. The tail to the bottom is a combination of the radiative tail and background from electrons striking the target ladder.}
    \label{fig:elastic_cluster}
\end{figure}

\begin{figure}
    \centering
    \includegraphics[width=\linewidth]{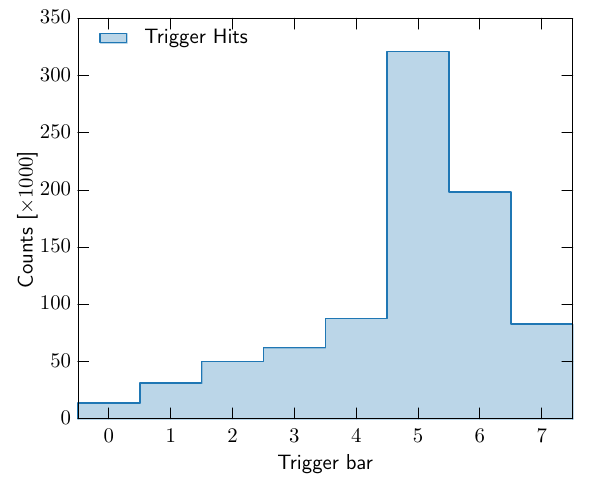}
    \caption{The distribution of hits in the trigger hodoscope.}
    \label{fig:trigger_hitmap}
\end{figure}

\begin{figure}
    \centering
    \includegraphics[width=\linewidth]{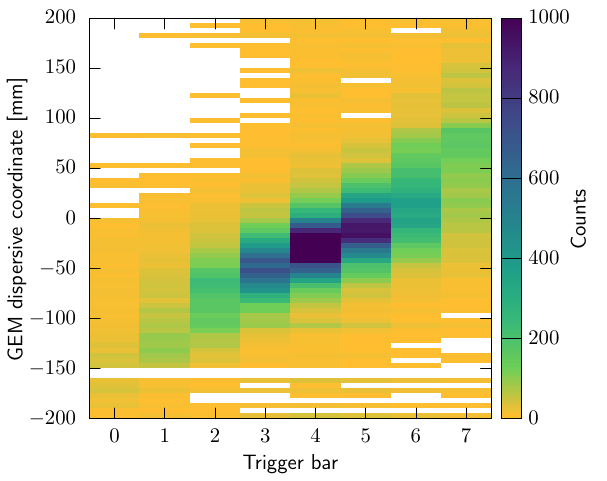}
    \caption{Correlation between the trigger bar with a hit in an event and the dispersive coordinate of the GEM cluster. The white stripe on the vertical coordinate at -150 mm is due to a dead region in the GEM.}
    \label{fig:GEM_trig_corr}
\end{figure}

The full detector stack for both spectrometers was operated for the majority of both data taking campaigns. During data-taking, GEM detectors would spontaneously develop sparking, rendering them inoperable. The detectors were replaced with on-site spares as needed. However for the last week of beam time, the top GEM on the 20$^{\circ}$ spectrometer failed, and there were no working replacements. As such, the data for 20, 25, and 30 MeV do not have two working GEMs on the right spectrometer. The trigger scintillators performed well during data taking. Final efficiency studies are underway but initial efforts indicated the trigger scintillators were approximately 70 \% efficient and the GEMs were approximately 30 \% efficient. The causes of the low efficiencies are unknown and further studies are ongoing.

To identify a clear signal, we first set the spectrometers to 10 MeV with a few-nA beam current. The elastic peak, as well as dead strips on the GEM, can be clearly seen in the Right Bottom GEM in Fig.~\ref{fig:elastic_cluster}. The corresponding trigger hitmap is shown in Fig.~\ref{fig:trigger_hitmap}, where the clear peak in bars five and six are indicative of elastic scattering.

Individual hits were identified in the GEM and trigger scintillators. To check correlations between detector components, the cluster coordinate of the Left Top GEM was plotted against the trigger bar with a hit in the event. A clear correlation between the detectors can be seen in Fig.~\ref{fig:GEM_trig_corr}.

To suppress background, tracks formed between the GEMs are projected onto the trigger paddles. The GEMs utilise multi-frame readout, providing some knowledge of the timing of the incident particle. By comparing the timing of the cluster with the time in the trigger scintillator, it is possible to reject background, noise, and accidental hits.

\subsection{Elastic Scattering from Carbon}

\begin{figure}
    \centering
    \includegraphics[width=\linewidth]{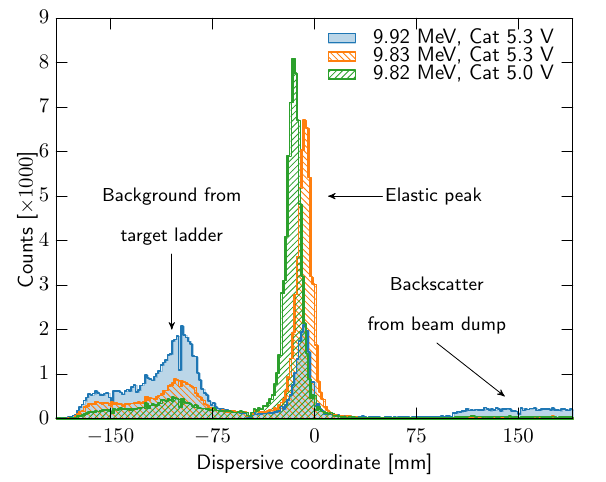}
    \caption{Distribution of hits in the dispersive coordinate of the right bottom GEM. The elastic peak can clearly be seen, as well as the background from the target ladder. It is clear that a cathode setting of 5.0~V (green) provides the best signal to noise ratio. The cathode settings of 5.3~V (orange and blue), show a worse signal to background. The difference between the orange and blue distributions arise by changes in tunes to the beam from the accelerator group.\label{fig:elastic_darkcurrent}}
\end{figure}

The first measurements taken during the commissioning runs were elastic scattering from carbon. This serves as a useful calibration point, and has a well known scattering rate for normalization to other measurements. 

As mentioned above, there was significant background due to the dark current present in the accelerator. After the improvements made to the operating conditions of the accelerator discussed above, a much cleaner elastic scattering spectrum could be seen.

The background due to the dark current could not be entirely removed from the elastic scattering datasets. The low-momentum electrons are over-steered by the final bending dipole and interact with the 1~cm thick aluminium target ladder. As the target ladder is much thicker than the target, even a small amount of dark current causes a large signal. Figure~\ref{fig:elastic_darkcurrent} shows the elastic peak as well as the background from the target ladder on different days and with different cathode settings. Nevertheless, the carbon peak is seen clearly enough that it can be used to calibrate the magnetic spectrometers, and test their linearity. Hysteresis effects in the magnets are still being studied, but initial results indicate such effects can be modeled in the simulation to high precision.

\subsection{M{\o}ller Scattering}

\begin{figure}
    \centering
    \includegraphics[width=\linewidth]{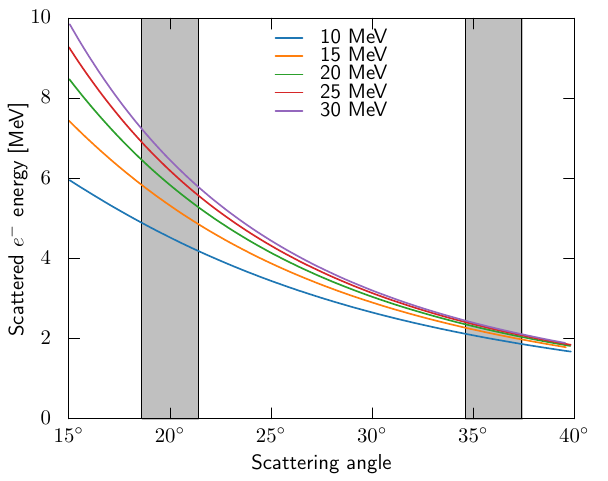}
    \caption{Energy of scattered M\o ller electrons as a function of scattering angle for different beam energies. From bottom to top the beam energies are 10, 15, 20, 25, and 30 MeV. The gray bands represent the angular acceptance of the two spectrometers.}
    \label{fig:moller_energies}
\end{figure}

\begin{figure}
    \centering
    \includegraphics[width=\linewidth]{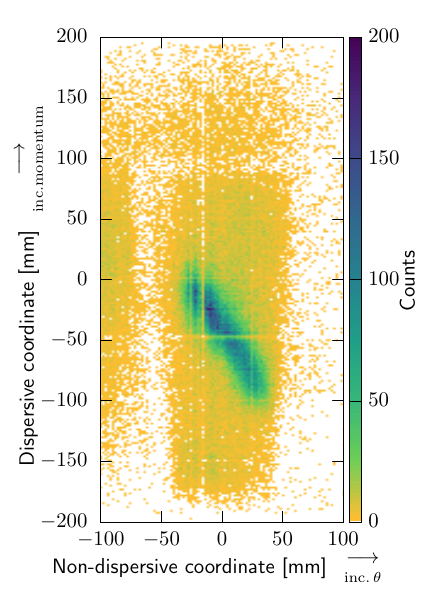}
    \caption{M\o ller Scattering from 10 MeV electrons scattering at 20$^{\circ}$ on a \SI{1}{\micro\metre} C target. The events on the left of the figure are background arising from the beam interacting with the beam pipe. The events on the top of the figure come from backscatter from the beam dump. The trapezoidal outline of the spectrometer exit flange can be clearly seen.}
    \label{fig:Moller_10MeV}
\end{figure}

A measurement of radiative M{\o}ller scattering is desirable as a test of QED radiative corrections.  Using the \SI{1}{\micro\metre} carbon target, and several beam energies, the spectrometers can be tuned to the central momentum for elastic M\o ller scattering. The M\o ller scattering energy at 36$^{\circ}$ does not change significantly, but at 20$^{\circ}$ the scattered energy varies to provide a more interesting measurement, see Fig.~\ref{fig:moller_energies}.

Substantial data were taken to measure radiative M\o ller scattering at all beam energies. While this data requires further analysis to produce a cross section, an initial yield plot can be seen in Fig.~\ref{fig:Moller_10MeV} for M\o ller scattering of 10 MeV electrons on the \SI{1}{\micro\metre} C target into the 20$^{\circ}$ spectrometer. The energy dependence of the cross section can clearly be seen. Data was taken using both the left and right spectrometers at all beam energies, while the non-M\o ller spectrometer was tuned to the corresponding elastic scattering energy to be used as a luminosity monitor.

This data will be compared to radiative event generators in future work. The aim will be to test radiative corrections in a regime where the electron mass is non-negligible compared to the beam energy.

\section{Conclusion}

The DarkLight experiment, located at the TRIUMF ARIEL e-linac, will search for a fifth force carrier in a model-independent way. Motivated in part by the X17 anomaly observed by the ATOMKI experiment, the experiment will ultimately target a search for a new boson with a mass of 17 MeV/$c^2$.
After completing installation in the summer of 2025, the experiment began its multi-step measurement program. 

An initial commissioning phase of the experiment was completed, along with its associated measurements, in early 2026 utilizing an electron beam energy of up to 30~MeV. Future running will be required to complete the commissioning of the experiment at the higher electron beam intensities demanded by the BSM physics search. In this paper we have documented the design and installation of the experiment, as well as the preliminary commissioning results, focusing on elastic $eC$ scattering and radiative M\o ller scattering. Future physics results will be the subject of upcoming papers.

In 2027, we anticipate that the high intensity commissioning of the experiment and the ARIEL electron beam will be completed and, subsequently, the 1000 h search will be performed using the existing 30~MeV electron beam energy. This search will exclude an area of phase space that corresponds to a boson with a mass of 13~MeV/$c^2$. Subsequently, the experiment will mount the full proposed search for a new boson with a mass of 17~MeV/$c^2$ after the accelerator has undergone an upgrade to increase the beam energy to 50~MeV. 

\section{Acknowledgements}

The DarkLight experiment is thankful for the support of the TRIUMF laboratory, especially the dedicated technical staff who assisted with the installation and commissioning of the experiment. We are also thankful for the engineering support from the MIT-Bates laboratory who were responsible for the mechanical design of the experiment. DarkLight is supported by the U.S. Department of Energy grant no.~DE-FG02-94ER40818, DE-SC0024464, and DE-SC0013941, by the U.S. National Science Foundation MRI PHY-1436680, by the MIT GMAF Palestine Fellowship Program, by the Stony Brook University URECA program, by the Canadian Natural Sciences and Engineering Research Council grant no. SAPEQ-2024-00004 and SAPPJ-2025-00043, the National Research Council of Canada, and by TRIUMF. We are also grateful for the student support provided by the TRIUMF Azuma Fellowship program, the International Atomic Energy Agency, the Government of Alberta, and the Canada First Research Excellence Fund via the Arthur B. McDonald Canadian Astroparticle Physics Research Institute.

\bibliographystyle{elsarticle-num-names}
\bibliography{bibliography}
\end{document}